\documentclass[journal=cmatex,manuscript=article,layout=twocolumn]{achemso}
\pdfoutput=1

\usepackage{graphicx}
\usepackage{epsfig}
\usepackage{subfigure}
\usepackage{color}
\usepackage{latexsym}
\usepackage{amsmath}
\usepackage{url}
\usepackage{bm} % bold math
\usepackage{amssymb} % symbols
\usepackage{lscape} % landscape for big tables
\usepackage{gensymb} % degree

\newcommand{\ilm}{Institut Lumi\`ere Mati\`ere, UMR5306 Universit\'e
  Lyon 1-CNRS, Universit\'e de Lyon, F-69622 Villeurbanne Cedex,
  France} 
\newcommand{\ub}{Department of Physics, Universit\"{a}t
  Basel, Klingelbergstr. 82, 4056 Basel, Switzerland}
\newcommand{\wuhan}{School of Physics and Technology, Wuhan
  University, Wuhan 430072, China} 
\newcommand{\jena}{Institut f\"ur
  Festk\"orpertheorie und -optik, Friedrich-Schiller-Universit\"at
  Jena and European Theoretical Spectroscopy Facility, Max-Wien-Platz
  1, 07743 Jena, Germany} 
\newcommand{\halle}{Institut f\"ur Physik,
  Martin-Luther-Universit\"at Halle-Wittenberg, D-06099 Halle,
  Germany}

\title{Identification of novel Cu, Ag, and Au ternary oxides from
  global structural prediction}

\author{Tiago F.T. Cerqueira}
\affiliation{\ilm}
\affiliation{\jena}
\author{Sun Lin}
\affiliation{\ilm}
\affiliation{\wuhan}
\author{Maximilian Amsler}
\affiliation{\ub}
\author{Stefan Goedecker}
\affiliation{\ub}
\author{Silvana Botti}
\affiliation{\ilm}
\affiliation{\jena}
\author{Miguel A.L. Marques}
\email{marques@tddft.org}
\affiliation{\ilm}
\affiliation{\halle}

\begin{document}

%\today
%\maketitle

\begin{tocentry}
\begin{center}
 \includegraphics[scale=0.15]{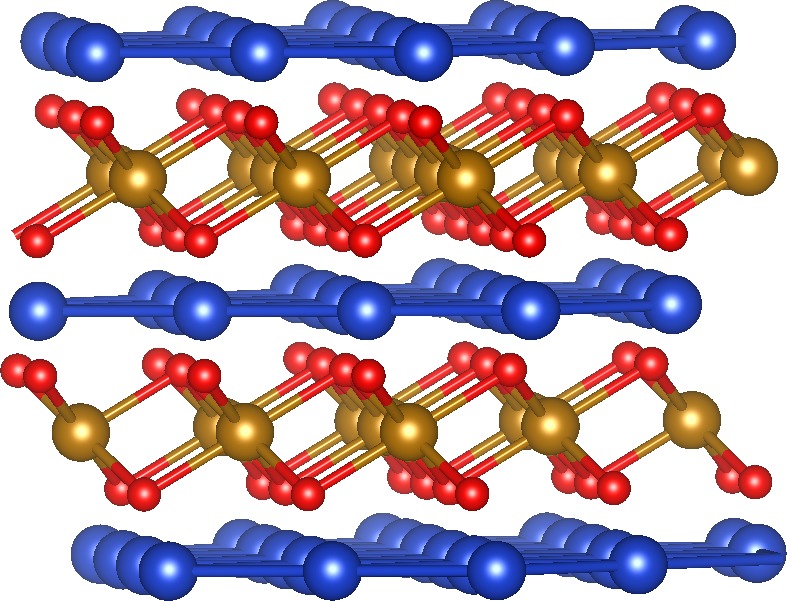} \\
 The delafossite crystal structure.
\end{center}
\end{tocentry}

\begin{abstract}
We use {\it ab initio} global structural prediction, and specifically
the minima hopping method, to explore the periodic table in search of
novel oxide phases. In total, we study 183 different compositions of
the form MXO$_2$, where M=(Cu, Ag, Au) and X is an element of the
periodic table. This set includes the well-known Cu delafossite
compounds that are, up to now, the best $p$-type transparent
conductive oxides known to mankind. Our calculations discover 81
stable compositions, out of which only 36 are included in available
databases. Some of these new phases are potentially good candidates
for transparent electrodes. These results demonstrate, on one hand,
how incomplete is still our knowledge of the phase-space of stable ternary
materials. On the other hand, we show that structural prediction
combined with high-throughput approaches is a powerful tool to extend
that knowledge, paving the way for the experimental discovery of new
materials on a large scale.
\end{abstract}

\section{Introduction}

Most of the knowledge painfully accumulated over centuries concerning
the crystal structure of (inorganic) materials is nowadays gathered in
generally available databases. The most used of these, the Inorganic
Crystal Structure Database (ICSD)~\cite{ICSD} contains around 170,000
entries. These entries include unfortunately duplicated structures,
insufficiently characterized phases (e.g., missing the positions of H
atoms), and several alloys. If we restrict ourselves to well-defined
crystal structures we find around 40,000 entries. 

%This number can be compared, e.g., to the number of chemical
%substances in the CAS registry (more than 88 million), to the number
%of macromolecules in the RCSB Protein Data Bank (more than 100,000),
%or even the number of crystals (organic and inorganic) in the
%Crystallography Open Database (around 300,000).

It is this amount of 40,000 entries that has been studied
theoretically, in a systematic manner, over the last few years using
high throughput techniques. The results of these large-scale studies
can be found in excellent publicly available databases, such as the
Materials Project~\cite{MaterialsProject}, the Open Quantum Materials
Database~\cite{Wolverton2013}, or the Ab-initio Electronic Structure
Library AFLOWLIB~\cite{Curtarolo2012227}. These are then used as a
starting point for sophisticated machine learning techniques that try
to design new materials with tailored properties. These techniques
have been touted as the most cost effective path to the discovery of
new materials in a diverse range of applications, such as Li
batteries~\cite{CederMRS,Ceder-chemmat2012},
thermoelectricity~\cite{CurtaroloPRX,Ophale2013,Carrete2014},
photovoltaics~\cite{Yu2012,Yu2013,Hautier2013,Hautier-CM2014}.

But let us go a step back, and look at the 40,000 materials we know
nowadays. The first question that comes to our mind is if this set is
representative of the number of (thermodynamically stable) materials
that we can create in a lab. The answer to this question is probably
no. It is true that elementary substances and binary compounds are
relatively well studied, but enormous gaps still exist in our
knowledge of ternary and multinary materials. Of course, a systematic
experimental endeavor of synthesis and characterization of all
possible phases is extremely expensive and time-consuming. It is for
this task that numerical simulations appear as the most cost-effective
way to explore the gigantic search space of compositions at our
disposal.

In this context, most theoretical studies follow a very simple
recipe. It is well-known that Nature often chooses the same solution
to similar problems. Therefore, one takes a known, experimentally
characterized structure, change its chemical composition, and hope
that this will also be a (dynamically and thermodynamically) stable
phase for the new composition. Sometimes the chemical composition can
be varied by ``brute force'' or one can use sophisticated machine
learning algorithms to predict what are the most promising
substitutions~\cite{Morgan2005,Hautier-IC2011,Fujimura-AEM2013,Pilania2013}. A
recent example of the first approach is the work of Carrete and coworkers~\cite{Carrete2014}, who computed all possible compositions of
half-Heusler compounds in the cubic structure (around 80,000
possibilities), while the latter approach was used, for example, by
Hautier and coworkers~\cite{Hautier-CM2010} who investigated Nature's
missing oxides, i.e., which oxides were thermodynamically stable, but
still unknown to mankind. It is evident that in either case the major
drawback of this approach is the impossibility to discover any
material with different crystal structure than those of the compounds
already contained in available databases.

In this Article, we address this problem and go a step further,
showing that structural prediction algorithms can be used on a large
scale together with high-throughput methods to predict the
lowest-energy crystal structures of materials with unreported chemical
compositions. The problem that global structural prediction methods
try to solve is very simple to explain: given the chemical composition
of a solid, obtain the minimum energy crystal structure, i.e. its
ground-state. In possession of this structure, we can then evaluate
its energy and a number of spectroscopic properties using standard
methods based on density functional theory and beyond. Of course, the
number of possible minima in a solid increases exponentially with the
size of the unit cell, which makes structural prediction an extremely
hard numerical problem. Fortunately, several smart algorithms for
global structural prediction appeared over the last decade, ranging
from random search to evolutionary
algorithms~\cite{Oganov2006,Curtarolo2005,Ma-PRB-2010,Pickard2011}.

Our method of choice is the minima hopping
method~\cite{MHM1,MHM2}. This is an efficient algorithm designed to
predict the low-energy crystal structures of a system given solely its
chemical composition.  At a given pressure, the enthalpy surface is
explored by performing consecutive short molecular dynamics escape
steps followed by local geometry relaxations, taking into account both
atomic and cell variables. The initial velocities for the molecular
dynamics trajectories are chosen approximately along soft mode
directions, allowing efficient escapes from local minima and aiming
towards lower energy valleys. Revisiting already known structures is
avoided by a feedback mechanism. The minima hopping method has been
used for structural prediction in a wide range of
materials~\cite{PhysRevLett.108.065501,PhysRevLett.110.135502,PhysRevLett.111.136101,MRS_comm},
including the dependence on pressure~\cite{PhysRevLett.108.117004} and
the exploration of binary phase diagram~\cite{NJP2013} with remarkable
results.

We use this machinery to study a subset of oxides, namely those having
a composition of the type (Cu,Ag,Au)XO$_2$, which includes the
``famous'' delafossites
Cu(Al,In,Ga)O$_2$~\cite{Kawazoe97,Yanagi01,Yanagi02,Ueda01}. These
compounds of this family are still the most promising p-type
transparent conductive oxides (TCO) known to mankind.  TCOs possess
the uncommon property of being at the same time transparent to the
visible spectrum and good electric conductors. These properties make
them indispensable for many high-technology devices which require
transparent contacts, such as flat panel displays, touch screens,
thin-film and stacked solar cells, functional windows, etc. Good
electron (n-)doped TCOs, namely those based on SnO$_2$, In$_2$O$_3$,
and ZnO, are already widely used in commercial applications. However,
potential p-type TCOs identified up to date have conductivities at
least one-two orders of magnitude lower than their n-type counterparts
and carrier mobilities too small for large-scale exploitation.  The
best p-type TCO until now is
CuCr$_{(1-x)}$Mg$_x$O$_2$~\cite{Nagarajan2001265}, which displays a
conductivity of 220 $\Omega^{-1}$cm$^{-1}$ and a hole mobility of
about 1 cm$^2$/Vs, while also suffering from poor transparency, with
transmission in the visible smaller than 30\%. The origin of the
relatively higher hole mobility of Cu delafossites relies on the fact
that the highest valence bands are obtained through the strong
hybridization of almost-degenerate oxygen $2p$ and copper $3d$
states. This hybridization reduces the localization of the top valence
states on oxygen atoms, leading to more dispersive $p$-$d$
anti-bonding bands with smaller hole effective masses\cite{Sheng2006}.
A tetrahedral coordination of oxide ions (as in delafossite crystals)
is particularly advantageous as it allows strong hybridization. Cu$^
{1+}$ (or equivalently Ag$^ {1+}$ and Au$^{1+}$) appear ideal elements
for creating a $p$-$d$ dispersive top valence while preserving
transparency, as their closed $d$ shell will prevent from absorption
in the visible. Even if it is believed that a Cu$^{+1}$ configuration
is the best to obtain TCOs, we decided nevertheless to explore blindly
also compositions that favor Cu$^{2+}$ or even Cu$^{3+}$, as we do not
want to preclude the possible formation of crystalline structures
different from delafossite that would let emerge interesting
electronic properties in different environments.

We can extract from the analysis above some expected good rules for
the design of improved p-type TCOs: (i) cations should have $d$ shells
proximate in energy to oxygen $2p$ states; (ii) cation $d$ shells
should be closed to avoid optical absorption in the visible range;
(iii) strong hybridization of oxygen $p$ and cation $d$ states is
required to increase the band dispersion\cite{Sheng2006}. Following
these simple ideas, few other Cu oxides were already successfully
tested in experiments after
Cu(Al,In,Ga)O$_2$~\cite{Kawazoe97,Yanagi01,Yanagi02,Ueda01}, such as
CuCrMgO$_2$~\cite{Nagarajan2001}, SrCu$_2$O$_2$~\cite{Kudo1998} or
LaCuOS~\cite{Ueda2000} and (Cu,Ag)ScO$_2$~\cite{Nagarajan2001265}. Our
aim is now to extend and make more systematic this investigation by
pre-screening {\it in silico} all possible Cu, Ag, and Au based
ternary oxides, in order to offer to experimentalists a reliable guide
on the stability and electronic properties of the still unknown
compositions. While we selected the broad family of (Cu,Ag,Au)XO$_2$
oxides as some of their members are examples of compounds satisfying the
empirical rules for good p-type TCO candidates, we decided at the same
time to limit as much as possible the restrictions imposed to the
choice of chemical elements inside this family, not to hinder the
emergence of unreported crystal structures with completely different
electronic properties.

Note that the set under consideration here is also a subset of the
systems studied using high-throughput techniques in the
above-mentioned work of Hautier and coworkers\cite{Hautier-CM2010}. As
a consequence, we will be able to provide a direct comparison of the
success rate of our mixed approach, that combines high-throughput with
structural prediction, in comparison with a purely high-throughput
study. For reference, we should keep in mind that three metastable
materials of the type (Cu,Ag,Au)XO$_2$ unreported in experimental
databases were obtained in Ref.~\citenum{Hautier-CM2010}, namely (i) a
trigonal phase of AgCoO$_2$ (50\,meV above the convex hull of
thermodynamic stability) based on the prototype experimental structure
of AgInO$_2$, (ii) a tetragonal phase of AgCsO$_2$ (340\,meV above the
hull) based on the prototype CuCsO$_2$ , and (iii) an hexagonal phase
of AgLaO$_2$ based on the prototype AgAlO$_2$ (4\,meV above the hull).

\section{Methodology}
\label{sec:methods}

We performed minima hopping simulations for all stoichiometries of the
type (Cu,Ag,Au)XO$_2$, where X is any element of the periodic table up
to Bi with the exclusion of the rare gases and the lanthanides. Forces
and energies were calculated within density functional
theory~\cite{Hohenberg1964,Kohn1965} in the projector augmented wave
(PAW) formalism~\cite{PAW} as implemented in {\sc
  vasp}~\cite{Kresse1996,PhysRevB.54.11169}. For a given
stoichiometry, the initial geometries were obtained randomly, ensuring
only that the minimal distance between the atoms was at least equal to
the sum of the covalent radii. The MHM searches were performed using
the Perdew-Burke-Erzernhof (PBE)~\cite{PhysRevLett.77.3865}
approximation to the exchange-correlation functional. We used default
``high'' accuracy energy cutoffs. Each minima hopping run was repeated
at least twice, using both one and two formula units (4 or 8
atoms). This may seem a relatively small number of atoms in the unit
cell, but if we look at the experimentally known structures of the
(Cu,Ag,Au)XO$_2$ family, we realize that almost all of them have a
ground-state with less than 8 atoms (exceptions are some metastable
phases of AgCO$_2$ with 16 atoms per unit cell and a stable phase of
AgBO$_2$ with 128 atoms per unit cell). Therefore, we assume that this
is not a major limitation.

We then compared the structures obtained in our runs with the ones
present in available experimental and theoretical
databases~\cite{MaterialsProject,Wolverton2013}. Almost all
experimental structures appeared at the early stages of our minima
hopping runs, which certainly proves the efficiency of the method for
this kind of task. Finally, we took the experimental structures and
other relevant theoretical phases that we discovered during our minima
hopping runs, and used them as prototypes for modified
stoichiometries. We believe that this procedure, which effectively
mixes high-throughput techniques with structural prediction is
essential in such large scale applications as each method can be used
to provide a check on the other.

In total we investigated 183 stoichiometries, and we obtained
$\sim$21,000 minima, of which $\sim$7,500 were further analyzed. In
this last step we followed the same protocol as in the Materials
Project database: spin-polarized calculation using the PBE~\cite{PhysRevLett.77.3865}
exchange-correlation functional, with the exception of the oxides of
Co, Cr, Fe, Mn, Mo, Ni, V, W where an on-site Coulomb repulsive
interaction U~\cite{LDA_U} with a value of 3.32, 3.7, 5.3, 3.9, 4,38,
6.2, 3.25, and 6.2\,eV, respectively, was added to correct the
$d$-states. PAW setups were taken from the version 5.2 of {\sc
  vasp}. At this stage the energy cutoff was set to 520\,eV
(irrespective of the elements considered) and k-point grids were
automatically chosen to ensure convergence to better than 2\,meV per
atom.  For all low-lying minima we studied the thermodynamic phase
equilibria of the ternary system, considering the energy balance with
respect to all possible decompositions in ternary, binary and
elementary compounds that respect the overall stoichiometry: i.e, we
measured the thermodynamic stability by calculating the energy
distance from the convex hull of stability, which is the set of lines
that connects the lowest energy ordered phases, using {\sc
  pymatgen}~\cite{pymatgen} and the data available in the Materials
Project database. According to this definition, a compound is stable
if its total energy distance to the convex hull is zero. Finally, the
crystallographic analysis of the structures was performed using {\sc
  findsym}~\cite{findsym}.

%%%%%%%%%%%%%%%%%%%%%%%%%
\begin{figure*}
  \includegraphics[width=1.99\columnwidth,angle=0,clip]{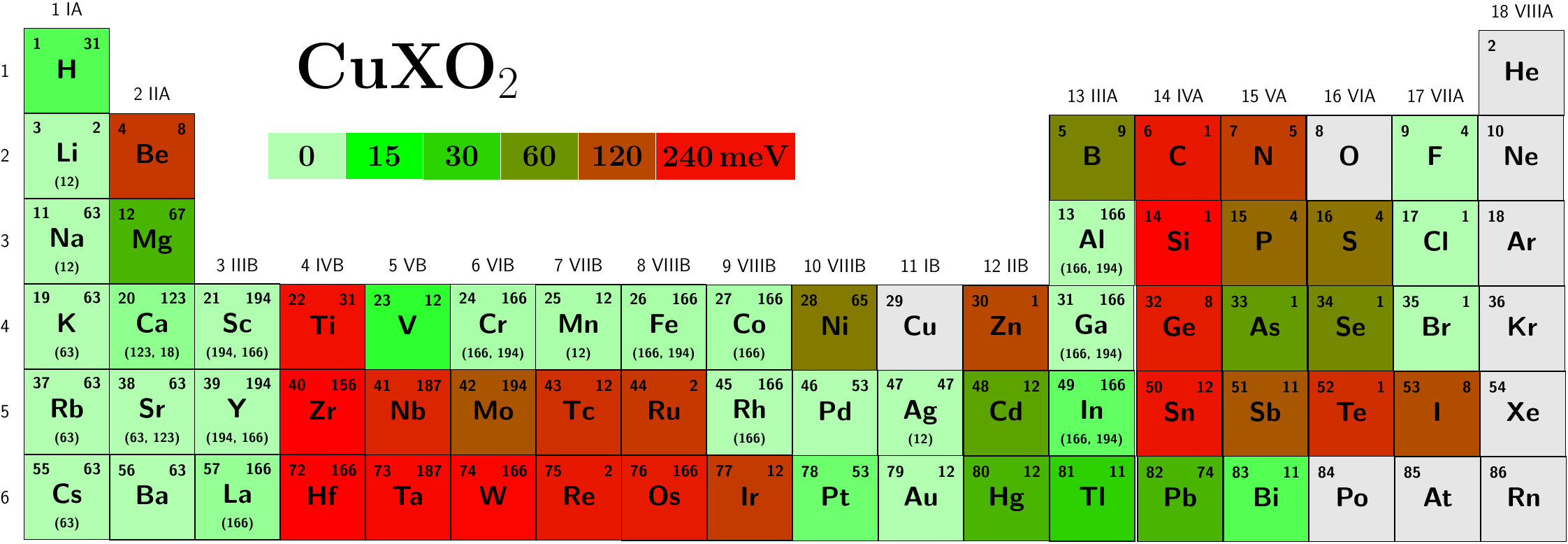}
  \includegraphics[width=1.99\columnwidth,angle=0,clip]{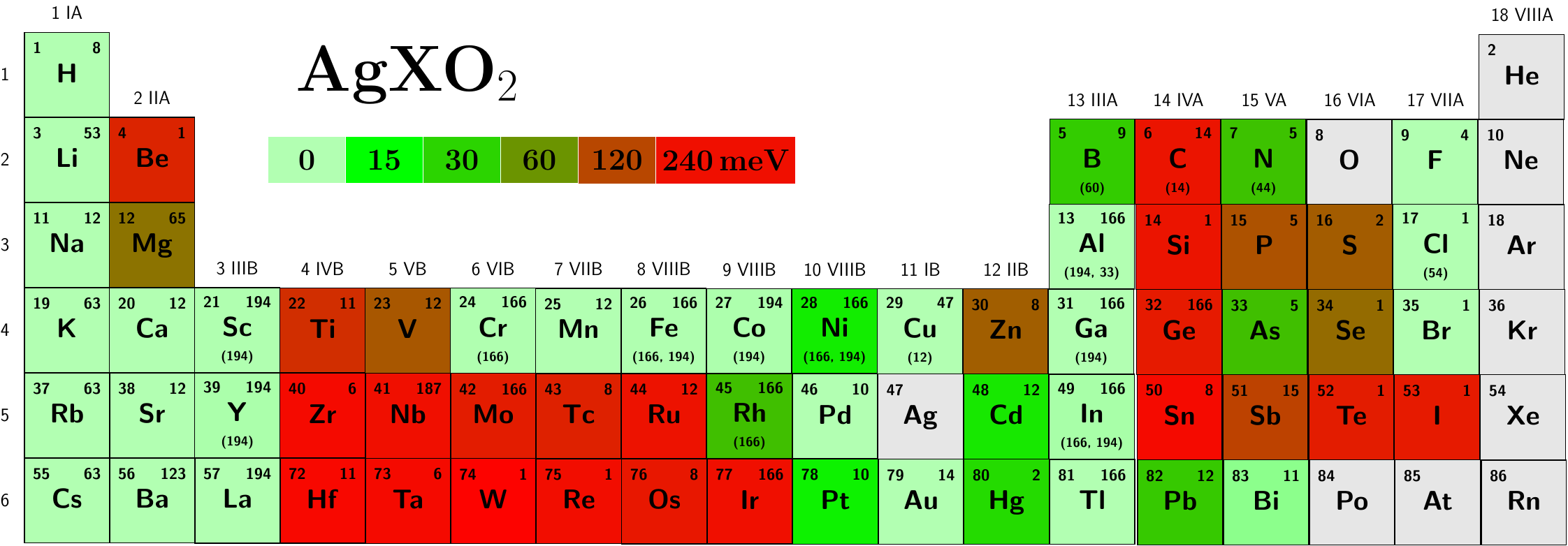}
  \includegraphics[width=1.99\columnwidth,angle=0,clip]{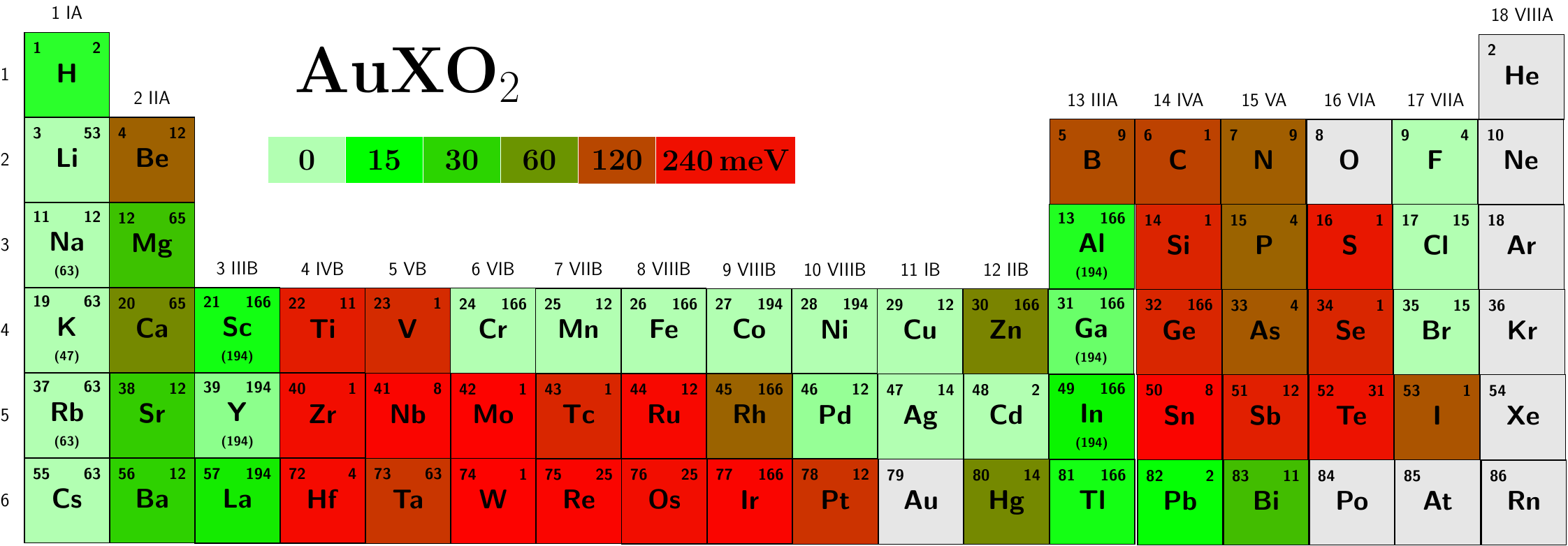}
  \caption{Distance to the convex hull (in meV per atom) for CuXO$_2$,
    AgXO$_2$, and AuXO$_2$. The colors indicate the distance to the
    convex hull of stability, light green meaning that the composition
    is thermodynamically stable. For each element we show the symbol
    (center), atomic number (top left), space group of the lowest
    energy structure (top right), and the space groups of the phases
    found in experimental databases (bottom, in parenthesis).}
  \label{fig:CuXO2}
\end{figure*}

\section{Results}

Our results are summarized in Fig.~\ref{fig:CuXO2}. In these periodic
tables the color scale indicates the distance of the calculated energy
of the ground-state structure to the convex hull of stability. Light
green cases indicate stable structures (i.e. that the specific
composition correspond to a crystal structure on the
hull). We further indicate the atomic number (upper left), the space
group we predict for the lowest energy phase (upper right), and the
space groups of the lowest-energy experimental structure that we found
in available databases (in parenthesis below the chemical
symbol). This means that all squares without a number below the
chemical symbol represent crystal structures predicted theoretically.
Further details can be found in the Supporting Information that
includes structural and electronic data of the most important
(meta-) stable structures.

As we can see from Fig.~\ref{fig:CuXO2}, many compositions are
thermodynamically stable, and many others are quite close to the
convex hull. Note that our calculations are performed at zero
temperature and pressure for perfect periodic crystals. However it is
well known that unstable phases can be stabilized by temperature,
pressure, defects, dopants, etc. Moreover, due to the theoretical
error associated to the calculation of the total energy, an inversion
of the ordering of the phases very close in energy is always possible.
In view of the above, we believe it is relevant to discuss all
compositions that are either thermodynamically stable or quasi-stable
(within 20\,meV from the convex hull). This choice of 20\,meV as a
threshold of stability comes from the observation that well studied
experimental compounds such as CuInO$_2$ and AgNiO$_2$ are above the
hull by 10-20\,meV in our calculations performed for perfect
stoichiometric bulk crystals. The set of structures that we select
within this threshold of 20\,meV contains therefore phases that have a
large chance of being synthesized experimentally. We find that this
stability condition is fulfilled by 81 compositions (30 containing Cu,
29 containing Ag, and 25 containing Au), of which only 36 (19
containing Cu, 11 containing Ag, and 8 containing Au) are present in
experimental databases.  We are now in the position to compare our
predictions with previous theoretical works.  We compared directly the
crystal structures whenever they were available for comparison. If the
structure in literature was not given, we used available information
on the symmetries and distance to the convex hull to compare with our
results. There are two phases (AgLaO$_2$ and AgCsO$_2$) previously
predicted in Ref.~\citenum{Hautier-CM2010}, while theoretical crystal
structures for CuBaO$_2$, CuAuO$_2$, CuHgO$_2$, AgLiO$_2$, AgNaO$_2$
can be found in the Materials Project database. In the two cases of
AgLaO$_2$ and AgNaO$_2$ we find the same crystal structure as
previously predicted. However, in the other five cases the MHM runs
were able to find lower-energy phases. In particular, the reported
structures of CuBaO$_2$, AgLiO$_2$, and AgCsO$_2$ were well above the
hull, while our lowest-energy structures are thermodynamically stable.
We should also observe that there may exist even lower energy
structures that were not detected in the MHM runs (because their unit
cell contains more than 8 atoms, for example). However, the compounds
we predict to be thermodynamically stable have a high chance of being
experimentally synthesizable, even if the actual atomic arrangement
may eventually differ from the one that we have determined. In this
sense, the periodic tables shown in Fig.~\ref{fig:CuXO2} offer a
simple guide to experimentalists by indicating which compositions are
expected to be easy and which are expected to be hard to synthesize,
based on thermodynamic considerations.

Before entering in a detailed discussion on the electronic properties
of the new stable phases found in our simulations, a few general
remarks are in order. (i)~There are some basic atomic arrangements
that are energetically favorable (with some exceptions) for many
compounds of the (Cu,Ag,Au)XO$_2$ family. The most common ones are the
delafossite structure of, e.g., CuFeO$_2$ (space groups 166 and 194)
and the tetragonal structure of CuCaO$_2$ (space group 123), together
with some slightly distorted, and therefore less symmetric,
variants. We stress, however, that some recurrent crystal structures
that we identified do not have any representative compound in
databases. (ii)~A majority of the structures are layered, but there is
a substantial variety in the geometry of the layers. (iii)~We observe
a mixture of semiconducting and metallic structures, and for some
compositions we can even find semiconducting and metallic phases
separated by only a few meV/atom.

\begin{figure}[t]
  \includegraphics[width=.99\columnwidth,angle=0]{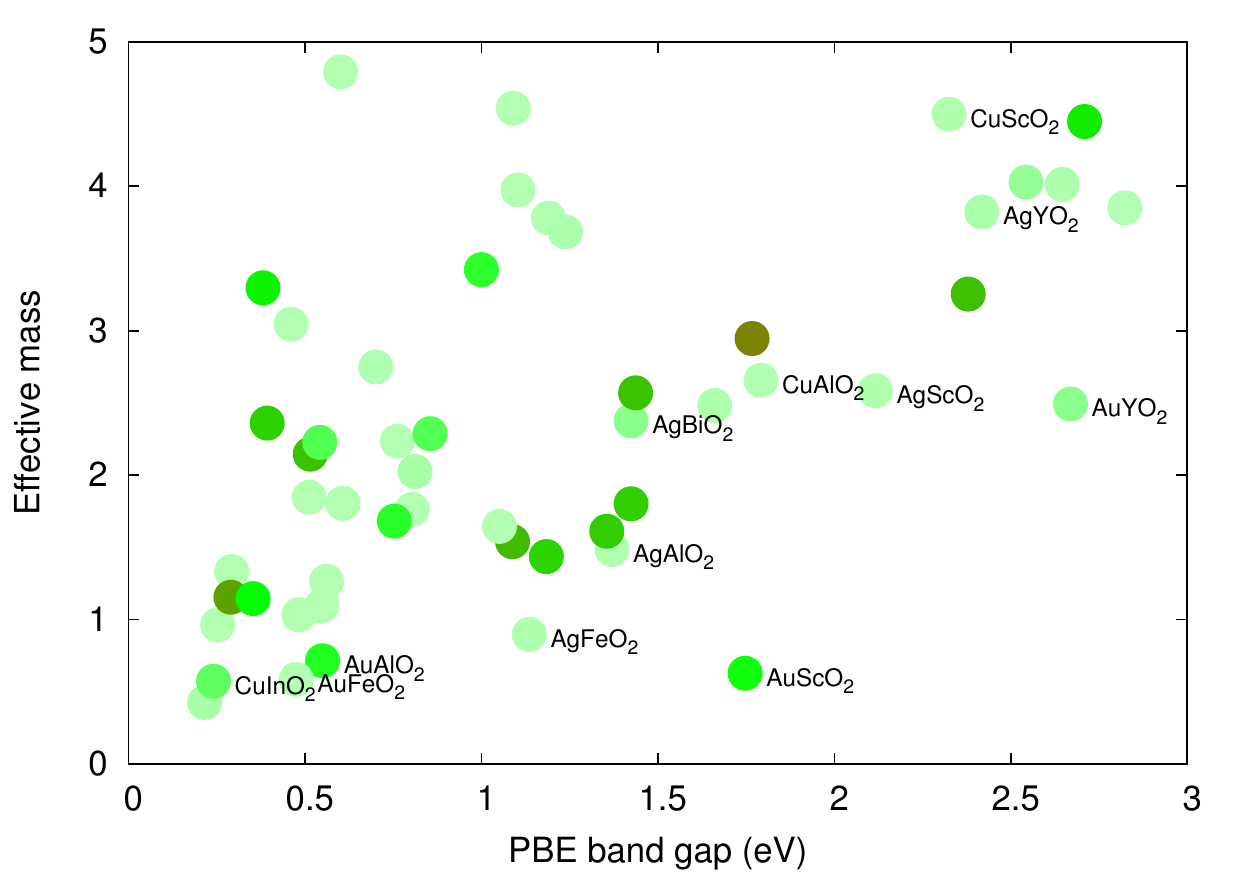}
  \caption{Effective hole masses as a function of the PBE(+U) band gap for
    all compounds that are within 70\,meV/atom from the convex
    hull. The color of the dots gives the distance to the hull and
    follows the same scale as in Fig.~\ref{fig:CuXO2}.}
  \label{fig:gapheffm}
\end{figure}

Having the crystal structures of new stable and metastable
(Cu,Ag,Au)XO$_2$ compounds opens the way for a series of further
theoretical studies. In fact, we can now calculate a wealth of
physical properties using the numerous methods and codes available for
theoretical spectroscopy. Our main motivation for the choice of this
class of systems is the p-type conductivity measured in Cu
delafossites. Therefore, we calculated the average hole effective mass
and the electronic band-gap for the ground-state of all the
stoichiometries studied. All calculations were perform with using {\sc
  pymatgen}~\cite{pymatgen} and {\sc BoltzTrap}~\cite{Boltztrap}
software packages. Following the same approach as
Ref.~\citenum{Hautier2013} we calculated the averaged hole effective
mass tensor for a carrier concentration of $10^{18}$~cm$^{-3}$ and a
temperature of 300~K. We then used the higher limit estimation for
$m^{*}_{h}$ (see Supplemental Information of
Ref.~\citenum{Hautier2013}). The results are summarized in
Fig.~\ref{fig:gapheffm}. In the figure, the color of the points
indicates thermodynamical stability, respecting the same color scale
as in Fig.~\ref{fig:CuXO2}.  We observe that the numbers we indicate
for the energy gap should be taken with care as they are Kohn-Sham
gaps obtained with the PBE(+U) exchange-correlation functional, and
are systematically
underestimated~\cite{PhysRevLett.104.136401,PhysRevB.82.085115}. The
true gaps are substantially larger (usually at least twice as large)
than the ones we indicate. Moreover, systems that are metallic within
PBE(+U) can sometimes be semiconducting in experiment, however the
inverse never happens. Note that band dispersions, and therefore
effective masses, are generally less sensitive to the choice of the
approximation used for the calculation of band structures.

For p-type transparent conductive materials we desire low-hole
effective masses (to improve conductivity), and high energy gaps (to
ensure transparency). Approximately in the center of
Fig.~\ref{fig:gapheffm} we find CuAlO$_2$, the compound where p-type
electrical conductivity in transparent thin-films was discovered for
the first time~\cite{Kawazoe97}. Its gap is sufficiently large (note
that the PBE gap of 1.8\, eV is substantially lower than the
experimental gap of around 3.5\,eV (see, e.g.,
Refs.\citenum{PhysRevLett.104.136401,PhysRevB.82.085115} and
references therein), but the hole effective mass is relatively
high. Indeed, the p-type conductivity in CuAlO$_2$ is still too low
for technological applications. From Fig.~\ref{fig:gapheffm} we can
see that several materials have the potential to outperform CuAlO$_2$,
such as AgScO$_2$, AuScO$_2$, AuYO$_2$, AgAlO$_2$, AgBiO$_2$, etc. Due
to the presence of (expensive) noble metals (Ag and Au), it is
unlikely that such materials can directly find large scale
applications in technology, nevertheless an experimental study of
conductivity in new interesting phases could give valuable ideas on if
(and eventually how) hole mobilities can be increased beyond present
limits.

In tables~\ref{tab:summary}, ~\ref{tab:summary2} and \ref{tab:summary3}  we summarize the information concerning the energy distances
from the convex hull, the band gaps and the hole effective masses.

\begin{table}[h]
\caption{\label{tab:summary}Space groups (Spg), distance to the convex hull
  (E$_{hull}$), gaps (Gap) and average hole effective mass ($m^{*}_{h}$) for
  semiconducting structures of the form CuXO$_2$ lying below $50$\,meV/atom from the convex
  hull.} 
\begin{tabular}[h]{r r r c r}
Structure & E$_{hull}$ & Spg & Gap & $m^{*}_{h}$ \\ \hline\\[-7pt]
 CuHO$_2$ &   8 &  31 & 0.5 &  2.23 \\
CuLiO$_2$ &   0 &   2 & 0.5 & 17.9 \\
 CuFO$_2$ &   0 &   4 & 0.8 & 11.8 \\
CuNaO$_2$ &   0 &  63 & 0.4 &  3.04 \\
CuMgO$_2$ &  43 &  67 & 0.0 &    -- \\
CuAlO$_2$ &   0 & 166 & 1.8 &  2.66 \\
CuClO$_2$ &   0 &   1 & 0.8 & 48.9 \\
 CuKO$_2$ &   0 &  63 & 0.8 & 13.7 \\
CuCaO$_2$ &   3 & 123 & 0.0 &    -- \\
CuScO$_2$ &   0 & 194 & 2.4 &  4.50 \\
 CuVO$_2$ &  11 &  12 & 1.0 &  3.42 \\
CuCrO$_2$ &   0 & 166 & 1.6 &  5.52 \\
CuMnO$_2$ &   0 &  12 & 0.1 &  0.06 \\
CuFeO$_2$ &   1 & 166 & 0.9 &  2.02 \\
CuCoO$_2$ &   0 & 166 & 1.1 &  5.63 \\
CuGaO$_2$ &   0 & 166 & 0.8 &  2.23 \\
CuBrO$_2$ &   0 &   1 & 0.9 &  9.67 \\
CuRbO$_2$ &   0 &  63 & 0.8 &  8.98 \\
CuSrO$_2$ &   0 &  63 & 0.0 &    -- \\
 CuYO$_2$ &   0 & 194 & 2.6 &  4.01 \\
CuRhO$_2$ &   0 & 166 & 0.7 &  2.75 \\
CuPdO$_2$ &   0 &  53 & 0.0 &    -- \\
CuAgO$_2$ &   0 &  47 & 0.0 &    -- \\
CuInO$_2$ &   7 & 166 & 0.3 &  0.57 \\
CuCsO$_2$ &   0 &  63 & 0.9 &  5.24 \\
CuBaO$_2$ &   0 &  63 & 0.0 &    -- \\
CuLaO$_2$ &   2 & 166 & 2.7 &  4.03 \\
CuYbO$_2$ &   0 & 123 & 0.0 &    -- \\
CuPtO$_2$ &   6 &  53 & 0.0 &    -- \\
CuAuO$_2$ &   0 &  12 & 0.0 &    -- \\
CuHgO$_2$ &  43 &  12 & 0.0 &    -- \\
CuTlO$_2$ &  31 &  11 & 0.4 &  2.36 \\
CuPbO$_2$ &  43 &  74 & 0.0 &    -- \\
CuBiO$_2$ &   8 &  11 & 1.0 &  2.29 \\
\end{tabular}
\end{table}

\begin{table}[h]
\caption{\label{tab:summary2}Space groups (Spg), distance to the convex hull
  (E$_{hull}$), gaps (Gap) and average hole effective mass ($m^{*}_{h}$) for
  semiconducting structures of the form AgXO$_2$ lying below $50$\,meV/atom from the convex
  hull.} 
\begin{tabular}[h]{r r r c r}
Structure & E$_{hull}$ & Spg & Gap & $m^{*}_{h}$ \\ \hline\\[-7pt]
 AgHO$_2$ &   0 &   8 & 0.0 &    -- \\
AgLiO$_2$ &   0 &  53 & 0.5 &  1.85 \\
 AgBO$_2$ &  33 &   9 & 1.4 &  1.61 \\
 AgNO$_2$ &  37 &   5 & 1.4 &  2.57 \\
 AgFO$_2$ &   0 &   4 & 0.6 &  7.61 \\
AgNaO$_2$ &   0 &  12 & 0.6 &  4.79 \\
AgAlO$_2$ &   0 & 166 & 1.4 &  1.49 \\
AgClO$_2$ &   0 &   1 & 0.6 & 2902 \\
 AgKO$_2$ &   0 &  63 & 0.9 &  5.80 \\
AgCaO$_2$ &   0 &  12 & 0.0 &    -- \\
AgScO$_2$ &   0 & 194 & 2.1 &  2.58 \\
AgCrO$_2$ &   0 & 166 & 1.7 &  2.48 \\
AgMnO$_2$ &   0 &  12 & 0.4 &  1.33 \\
AgFeO$_2$ &   0 & 166 & 1.1 &  0.90 \\
AgCoO$_2$ &   0 & 194 & 1.2 &  3.68 \\
AgNiO$_2$ &  20 & 166 & 0.0 &    -- \\
%AgCuO$_2$ &   0 &  47 & 0.0 &    -- \\
AgGaO$_2$ &   0 & 166 & 0.6 &  1.09 \\
AgAsO$_2$ &  38 &   5 & 2.4 &  3.25 \\
AgBrO$_2$ &   0 &   1 & 0.7 & 450 \\
AgRbO$_2$ &   0 &  63 & 1.0 &  8.80 \\
AgSrO$_2$ &   0 &  12 & 0.6 &  1.27 \\
 AgYO$_2$ &   0 & 194 & 2.4 &  3.82 \\
AgRhO$_2$ &  38 & 166 & 0.5 &  2.14 \\
AgPdO$_2$ &   0 &  10 & 0.1 &  0.03 \\
AgCdO$_2$ &  23 &  12 & 0.0 &    -- \\
AgInO$_2$ &   0 & 166 & 0.2 &  0.43 \\
AgCsO$_2$ &   0 &  63 & 1.1 &  4.54 \\
AgBaO$_2$ &   0 & 123 & 0.0 &    -- \\
AgLaO$_2$ &   0 & 194 & 2.8 &  3.85 \\
AgPtO$_2$ &  19 &  10 & 0.3 &  3.30 \\
AgAuO$_2$ &   0 &  14 & 0.6 &  1.80 \\
AgHgO$_2$ &  27 &   2 & 0.0 &    -- \\
AgTlO$_2$ &   0 & 166 & 0.0 &    -- \\
AgPbO$_2$ &  34 &  12 & 0.3 &  0.03 \\
AgBiO$_2$ &   3 &  11 & 1.4 &  2.38 \\
\end{tabular}
\end{table}
\begin{table}[h]
\caption{\label{tab:summary3}Space groups (Spg), distance to the convex hull
  (E$_{hull}$), gaps (Gap) and average hole effective mass ($m^{*}_{h}$) for 
  semiconducting structures of the form AuXO$_2$ lying below $50$\,meV/atom from the convex
  hull.} 
\begin{tabular}[h]{r r r c r}
Structure & E$_{hull}$ & Spg & Gap & $m^{*}_{h}$ \\ \hline\\[-7pt]
 AuHO$_2$ &  12 &   2 & 0.7 &  1.68 \\
AuLiO$_2$ &   0 &  53 & 1.1 &  3.78 \\
 AuFO$_2$ &   0 &   4 & 1.1 &  3.97 \\
AuNaO$_2$ &   0 &  12 & 1.1 &  7.72 \\
AuMgO$_2$ &  37 &  65 & 0.0 &    -- \\
AuAlO$_2$ &  12 & 166 & 0.6 &  0.71 \\
AuClO$_2$ &   0 &  15 & 0.3 &  0.96 \\
 AuKO$_2$ &   0 &  63 & 1.4 &  9.20 \\
AuScO$_2$ &  14 & 166 & 1.8 &  0.63 \\
AuCrO$_2$ &   0 & 166 & 1.2 &  1.65 \\
AuMnO$_2$ &   0 &  12 & 0.0 &    -- \\
AuFeO$_2$ &   0 & 166 & 0.6 &  0.59 \\
AuCoO$_2$ &   0 & 194 & 0.5 &  1.03 \\
AuNiO$_2$ &   0 & 194 & 0.0 &    -- \\
%AuCuO$_2$ &   0 &  12 & 0.0 &    -- \\
AuGaO$_2$ &   6 & 166 & 0.0 &    -- \\
AuBrO$_2$ &   0 &  15 & 0.0 &    -- \\
AuRbO$_2$ &   0 &  63 & 1.4 &  8.45 \\
AuSrO$_2$ &  33 &  12 & 1.3 &  1.80 \\
 AuYO$_2$ &   3 & 194 & 2.7 &  2.49 \\
AuPdO$_2$ &   2 &  12 & 0.0 &    -- \\
%AuAgO$_2$ &   0 &  14 & 0.6 &  1.80 \\
AuCdO$_2$ &   0 &   2 & 0.9 &  1.76 \\
AuInO$_2$ &  18 & 166 & 0.0 &    -- \\
AuCsO$_2$ &   0 &  63 & 1.5 &  6.59 \\
AuBaO$_2$ &  31 &  12 & 1.2 &  1.44 \\
AuLaO$_2$ &  21 & 194 & 2.7 &  4.45 \\
AuTlO$_2$ &  13 & 166 & 0.0 &    -- \\
AuPbO$_2$ &  15 &   2 & 0.4 &  1.14 \\
AuBiO$_2$ &  40 &  11 & 1.3 &  1.54 \\
\end{tabular}
\end{table}

These results call for more detailed
experimental studies of this restricted set of compounds, that should
be accompanied by more accurate calculations of the electronic band
gaps (i.e. using GW approaches beyond standard density functional
theory~\cite{PhysRevLett.104.136401,PhysRevB.82.085115}) and of
possible defects/impurities for p-type dopability. From a
purely theoretical point of view, the results of
Fig.~\ref{fig:gapheffm} demonstrate that, with our approach, we are
now able to go all the way from a simple stoichiometry to the
estimate of relevant material properties for a vast class of
materials.

We will now analyze more in detail the structural and electronic
properties of the different subclasses of compounds that we
identified. We will extend this discussion to phases that are closer
than 50 meV to the convex hull.

\subsection{Delafossites}
%%%%%%%%%%%%%%%%%%%%%%%%%
\begin{figure}[t]
  \includegraphics[width=.6\columnwidth,angle=0]{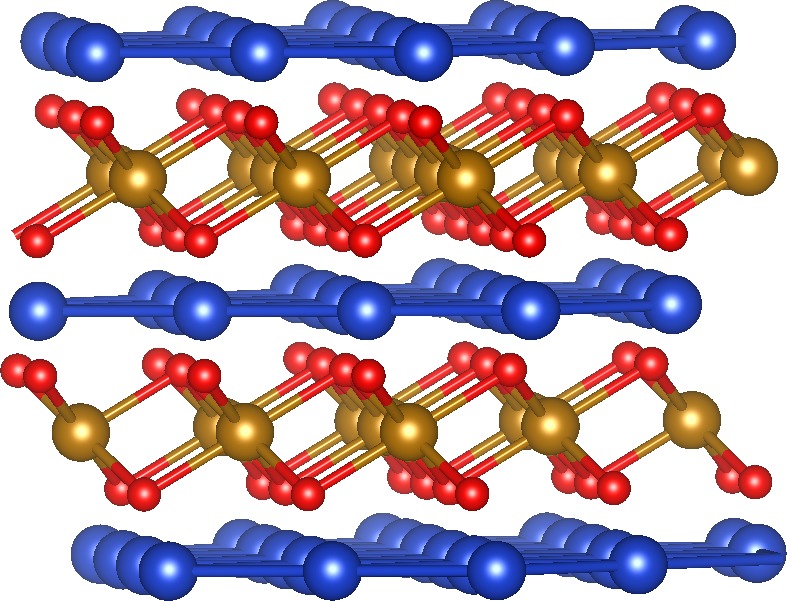}
  \caption{The delafossite structure (space group 166).}
  \label{fig:spg166}
\end{figure}

The trigonal (space group 166) or hexagonal (space group 194)
delafossite structures (depending on the stacking sequence) are the
crystal structure assumed by all stable compounds with X belonging to
group IIIA (Al, Ga, In), IIIB (Sc, Y, La), or X=Cr, Fe, Co, Ni, and
Rh. Also AgTlO$_2$ and AuTlO$_2$ (but not CuTlO$_2$) present a
low-energy delafossite structure. This phase is characterized by
XO$_2$ planes separated by flat hexagonal (Cu,Ag,Au) planes. The
energy differences between the trigonal and hexagonal phases is always
very small (of the order of few meV/atom), sometimes smaller than
the precision in our calculations. Thermodynamically stable
delafossite structures not present in databases are
AgTlO$_2$, AuCrO$_2$, AuFeO$_2$, AuCoO$_2$, AuNiO$_2$, AuTlO$_2$ and
AuLaO$_2$.

\subsection{Group IA}

%%%%%%%%%%%%%%%%%%%%%%%%%
\paragraph{Hydrogen}

\begin{figure}[t]
  \centering
  \begin{tabular}{cc}
    {\bf (a)} & {\bf (b)} \\
    \includegraphics[width=.49\columnwidth,angle=0]{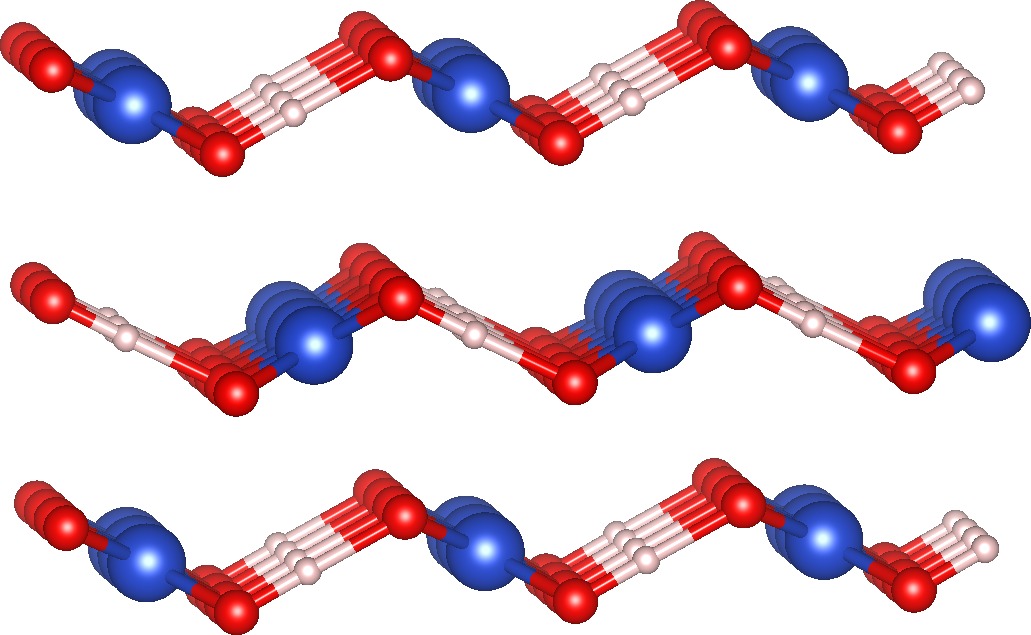} &
    \includegraphics[width=.49\columnwidth,angle=0]{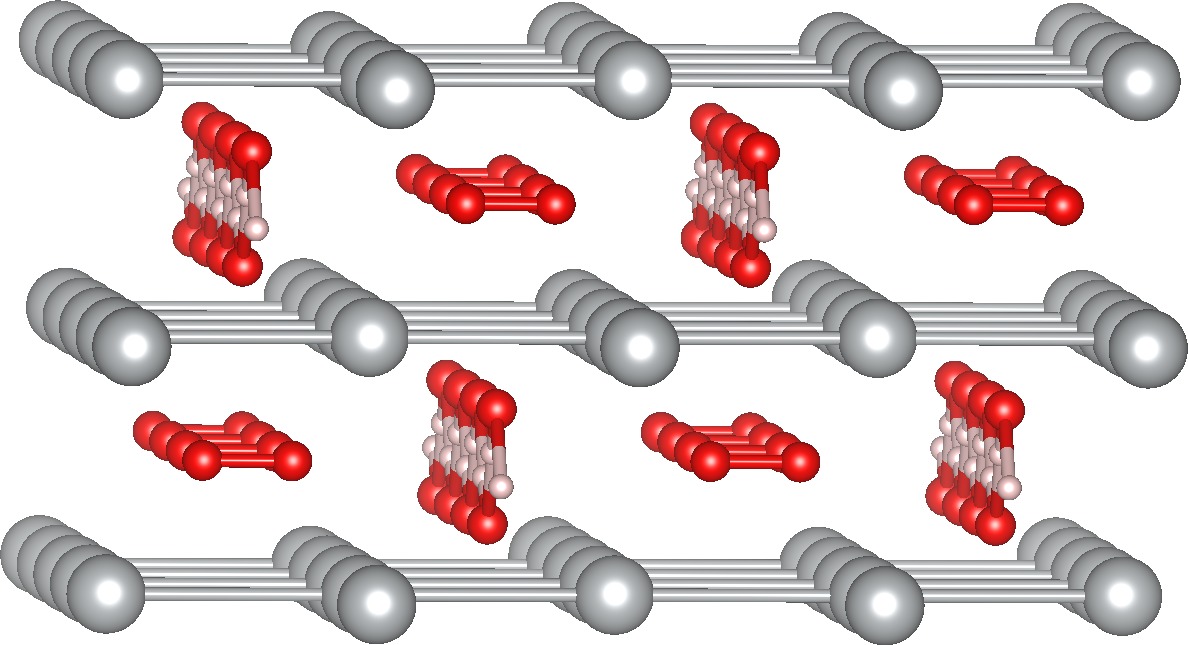}
  \end{tabular}

  {\bf (c)}

  \includegraphics[width=.49\columnwidth,angle=0]{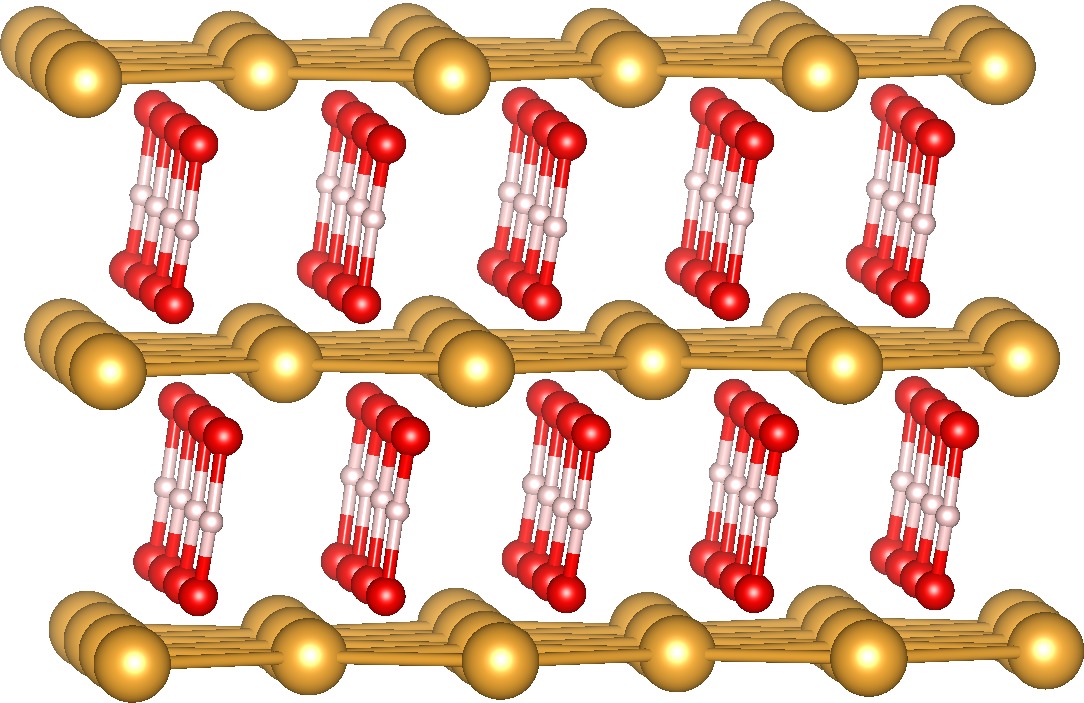}  
  \caption{The structures of (a)~CuHO$_2$ (space group 31),
    (b)~AgHO$_2$ (space group 8), and (c)~AuHO$_2$ (space group 2).}
  \label{fig:CuHO2}
\end{figure}

We found a series of distinct structures for (Cu,Ag,Au)HO$_2$ (see
Fig.~\ref{fig:CuHO2}). The Cu compound crystallizes in a layered
structure (space group 31) composed of zigzag layers of CuHO$_2$. It
is just above the convex hull at 8\,meV/atom, and is an indirect-gap
semiconductor with a PBE gap of 0.5\,eV. The Ag compound shows flat
layers of Ag (in a square lattice) separated by zigzag lines of OH and
O$_2$ dimers. This is a monoclinic (space group 8), metallic,
thermodynamically stable phase. Finally, AuHO$_2$ is a low-symmetry
structure (space group 2), composed by flat Au layers separated by
aligned HO$_2$ units. It is 11\,meV/atom above the convex hull and it
is an indirect gap semiconductor with a PBE gap of 0.7\,eV.

%%%%%%%%%%%%%%%%%%%%%%%%%
\begin{figure}[t]
  \centering
  \begin{tabular}{cc}
    {\bf (a)} & {\bf (b)} \\
    \includegraphics[width=.47\columnwidth,angle=0]{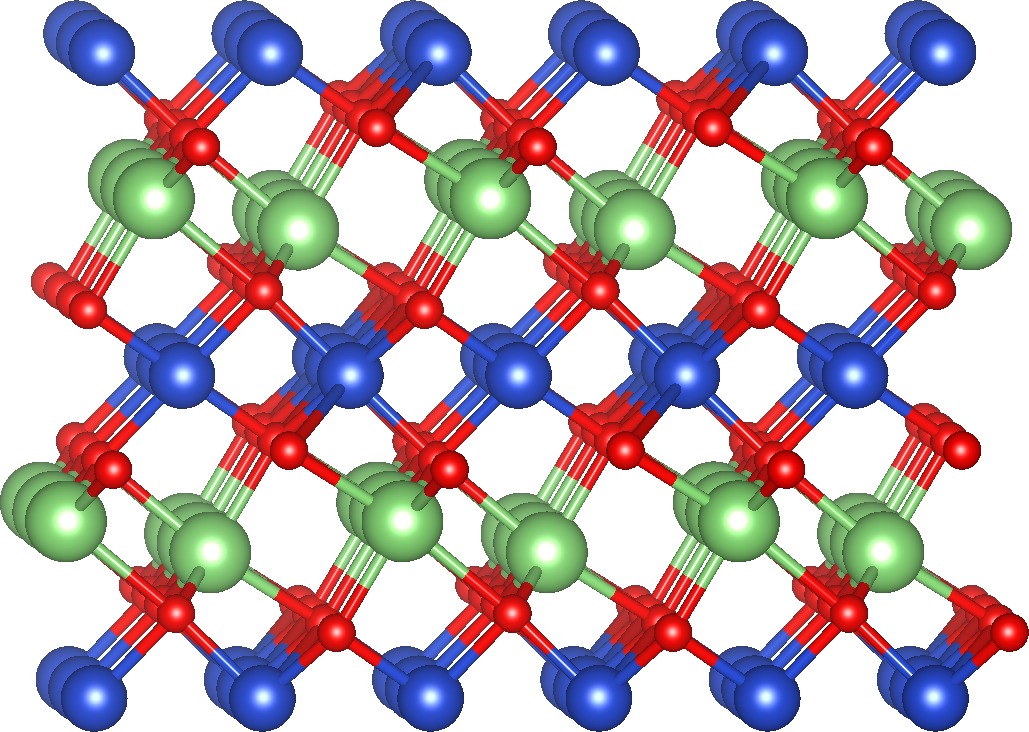} &
    \includegraphics[width=.51\columnwidth,angle=0]{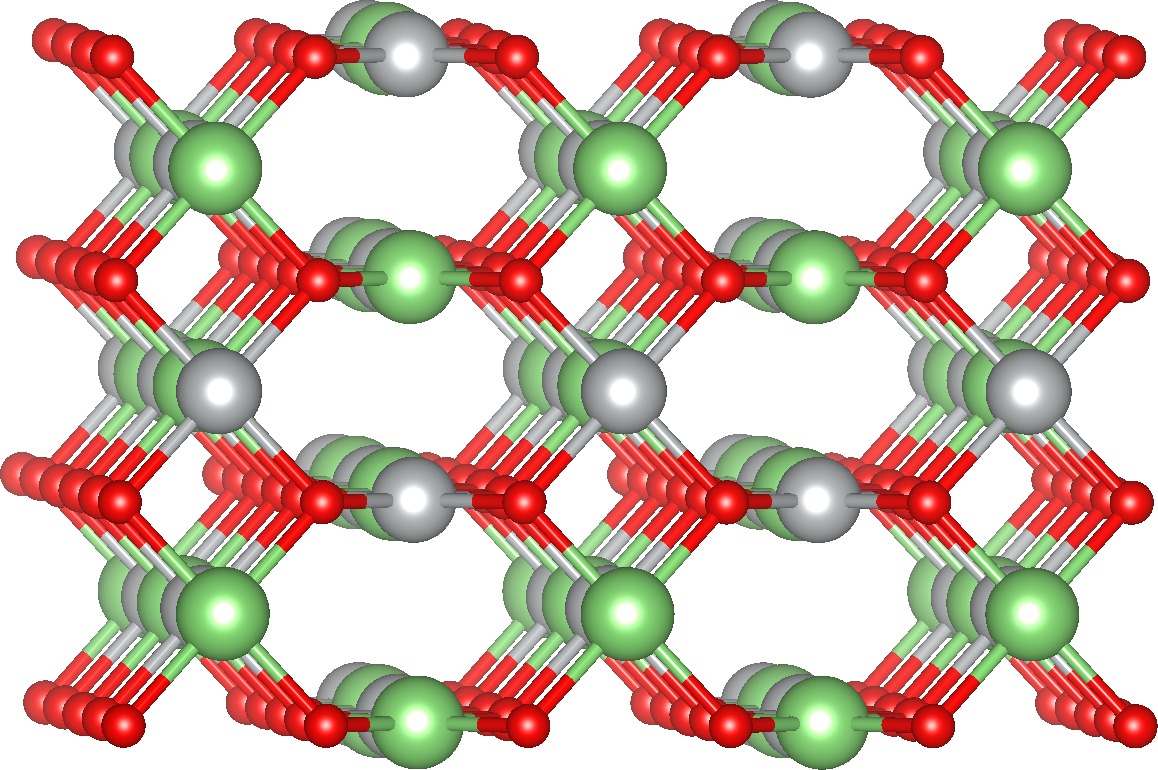}
  \end{tabular}
  \caption{The structures of (a)~CuLiO$_2$ (space group 2) and
    (b)~(Ag,Au)LiO$_2$ (space group 53).}
  \label{fig:CuLiO2}
\end{figure}

\paragraph{Lithium}

The structures of (Cu,Ag,Au)LiO$_2$ are composed of lines of
alternating (Cu,Ag,Au) and Li bonded by O in a three-dimensional
arrangement (see Fig.~\ref{fig:CuLiO2}). The Ag and Au compounds
crystallize in an orthorhombic structure (space group 53). A distortion
of this structure leads to the monoclinic phase (space group 12) found in
the databases for CuLiO$_2$, and a large distortion of this structure
reduces the symmetry to the space group 2 that we find as the
ground-state of CuLiO$_2$.  All compounds are indirect band-gap
semiconductors, with PBE gaps of 0.5\,eV (Cu), 0.5\,eV (Ag), and
1.1\,eV (Au).

%%%%%%%%%%%%%%%%%%%%%%%%%
\paragraph{Sodium}

\begin{figure}[t]
  \includegraphics[width=.49\columnwidth,angle=0]{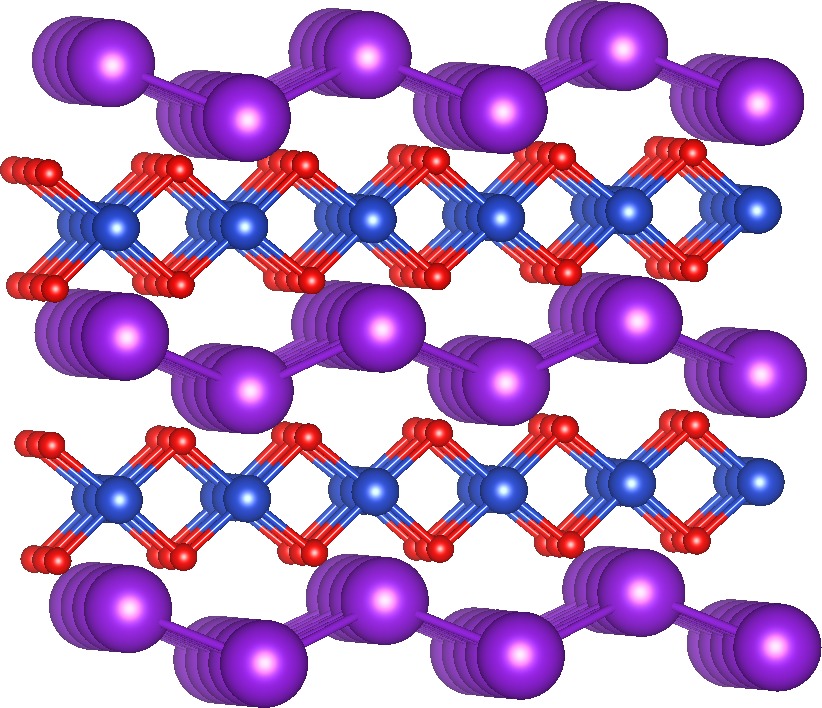}
  \caption{The structure of (Cu,Ag,Au)(K,Rb,Cs)O$_2$ (space group
    63).}
  \label{fig:spg63b}
\end{figure}

The compounds containing Na result in two structures very close in
energy and competing for the ground-state: the orthorhombic structure
(see Fig.~\ref{fig:spg63b}) that is the structure of the ternary
oxides containing K, Rb, and Cs, and the monoclinic phase that is a
slightly distorted version of the structure of (Ag, Au)LiO$_2$. For
CuNaO$_2$ we find that the former is the ground state (a mere 7\,meV
per atom below the monoclinic structure present in databases), while
for Ag and Au the latter is more stable. These compounds are all indirect-gap
semiconductors with PBE gaps of 0.4\,eV (Cu), 0.6\,eV (Ag), and
1.1\,eV (Au).

%%%%%%%%%%%%%%%%%%%%%%%%%

\paragraph{Potassium, Rubidium, Cesium}

The compounds of the form (Cu,Ag,Au)(K,Rb,Cs)O$_2$ crystallize in an
orthorhombic structure (space group 63), characterized by corrugated
square planes of (K,Rb,Cs) intercalated with flat stripes of CuO$_2$
where each Cu atom shares four O atoms with two neighboring Cu (see
Fig.~\ref{fig:spg63b}). All these materials are semiconducting with an
indirect PBE gap between 0.8 and 1.5\,eV that increases with the
size of the atoms from Cu to Au and from K to Cs.  Note that this is
also the lowest energy structure we found for CuNaO$_2$.

\subsection{Group IIA}

%%%%%%%%%%%%%%%%%%%%%%%%%
\paragraph{Magnesium}
\begin{figure}[t]
  \centering
  \begin{tabular}{cc}
    {\bf (a)} & {\bf (b)} \\
    \includegraphics[width=.42\columnwidth,angle=0]{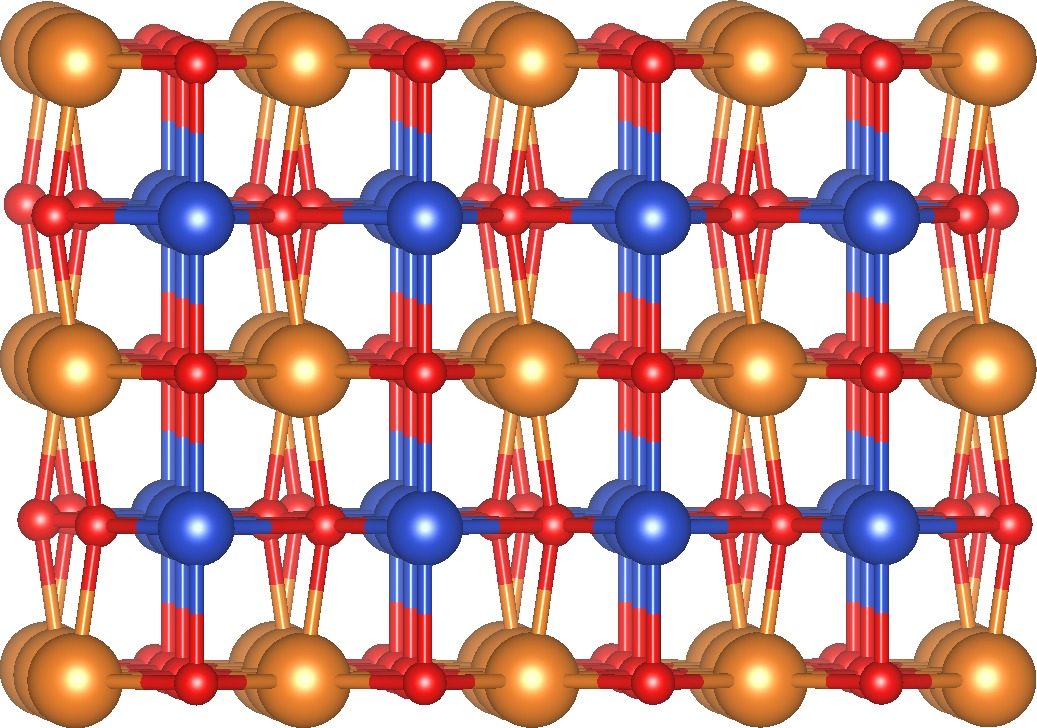} &
    \includegraphics[width=.56\columnwidth,angle=0]{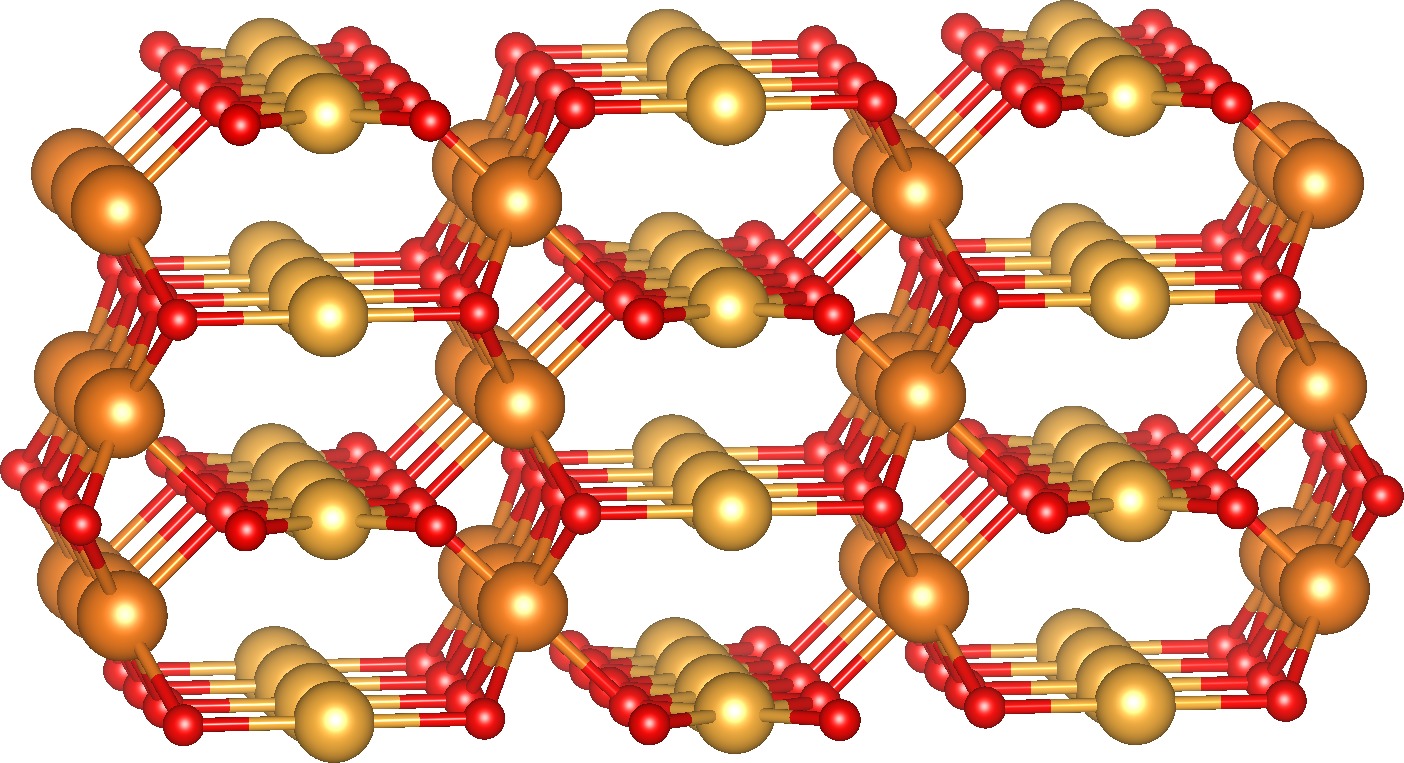}
  \end{tabular}
  \caption{The structures of (a)~CuMgO$_2$ (space group 67) and
    (b)~AuMgO$_2$ (space group 65)}
  \label{fig:CuMgO2}
\end{figure}

\begin{figure}[t]
  \centering
  \begin{tabular}{cc}
    {\bf (a)} & {\bf (b)} \\  
    \includegraphics[width=.49\columnwidth,angle=0]{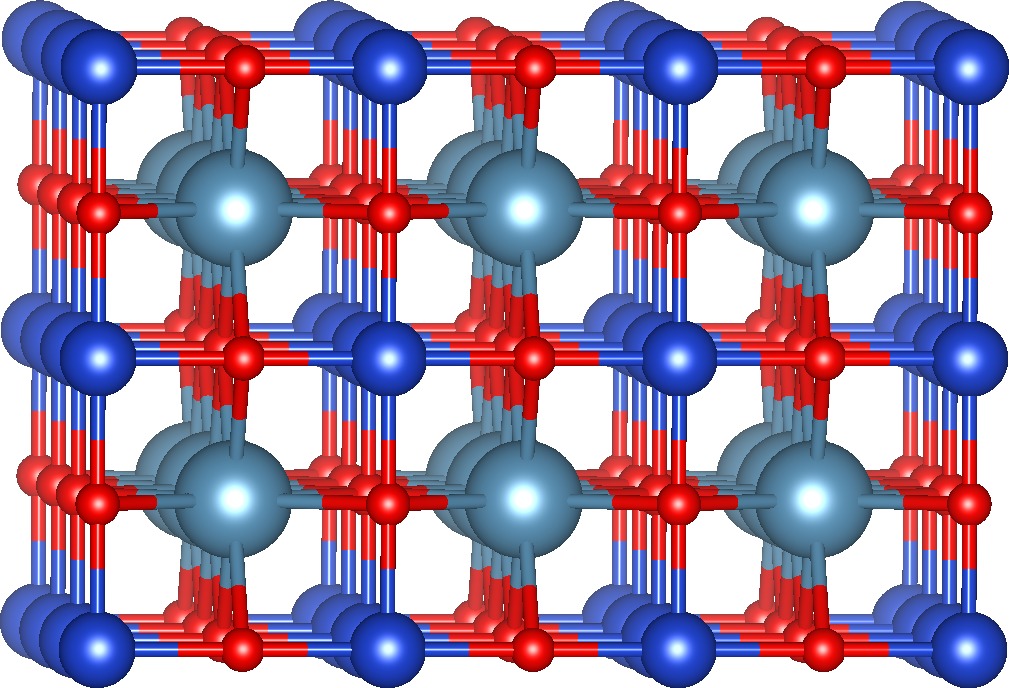} &
    \includegraphics[width=.45\columnwidth,angle=0]{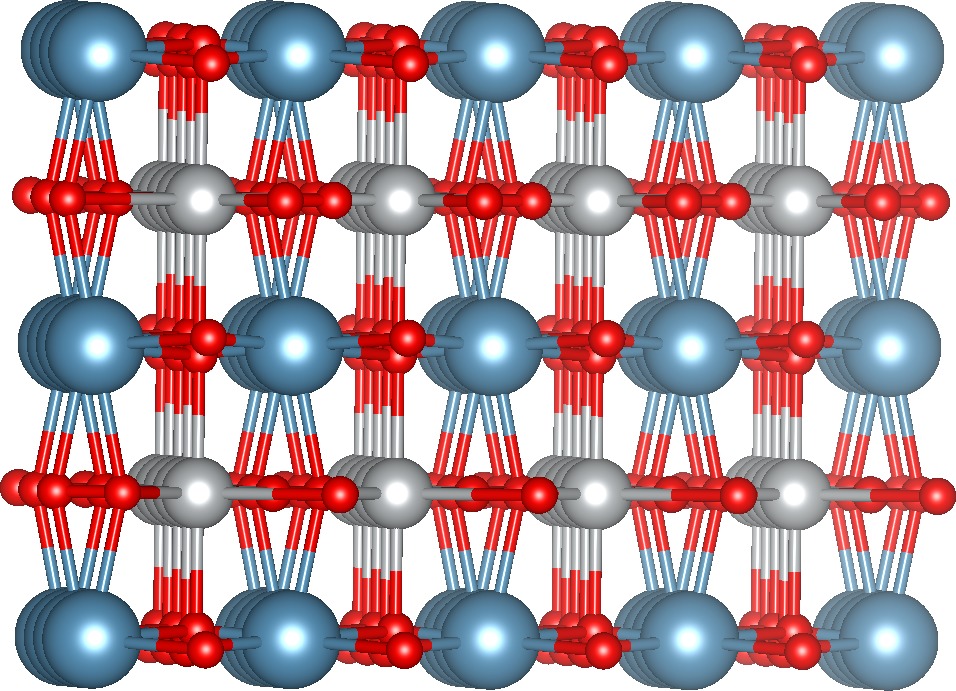}
  \end{tabular}
  
  {\bf (c)}

  \includegraphics[width=.49\columnwidth,angle=0]{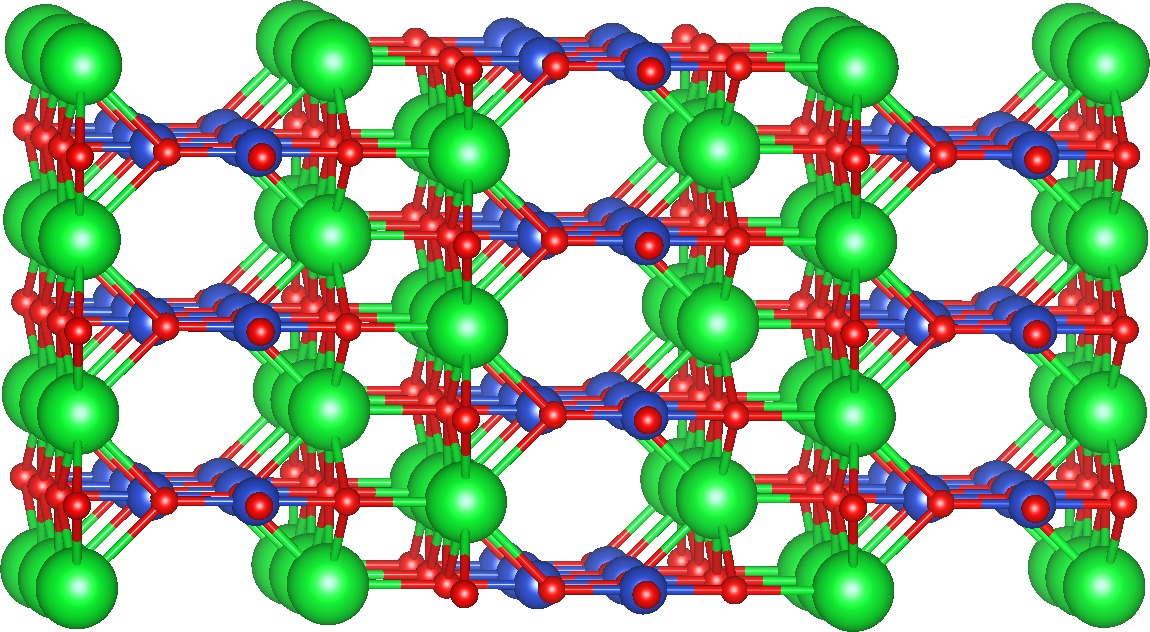}
  \caption{The structures of (a)~CuCaO$_2$ (space group 123),
    (b)~AgCaO$_2$ (space group 12), and (c)~CuSrO$_2$ (space group
    63).}
  \label{fig:CuCaO2}
\end{figure}

The element Mg is contained in two ternary oxides with
Cu (43\,meV/atom from the convex hull) and with Au
(36\,meV/atom). They are both orthorhombic, but quite dissimilar: the
structure CuMgO$_2$ (space group 67) is a distorted variation of the
tetragonal unit-cell (space group 123, see Fig.~\ref{fig:CuCaO2}) of,
e.g., CuCaO$_2$. AuMgO$_2$ is more complicated: its crystal
structure (space group 65) is similar to the one of (Ag,Au)LiO$_2$
(see right panel of Fig.~\ref{fig:CuLiO2}). Both structures are
metallic.

%%%%%%%%%%%%%%%%%%%%%%%%%
\paragraph{Calcium, Strontium, Barium}

Most of the (Cu,Ag,Au)(Ca,Sr,Ba)O$_2$ ternaries crystallize in a
tetragonal lattice with space group 123 (CuCaO$_2$, AgBaO$_2$) or in a
distorted version with a monoclinic space group 12 (AgCaO$_2$,
AgSrO$_2$, AuSrO$_2$, AuBaO$_2$). These two lattices are represented
in the top panels of Fig.~\ref{fig:CuCaO2}. On the other hand,
CuSrO$_2$ and CuBaO$_2$ have an orthorhombic structure (space group
63) that resembles to some to extent the one of AuMgO$_2$. Most of these
materials are metallic, with the exception of the ones that
crystallize in the monoclinic phase: AgSrO$_2$ has a PBE gap of
0.6\,eV, AuSrO$_2$ of 1.4\,eV and AuBaO$_2$ of 1.2\,eV. AgCaO$_2$
has zero PBE gap, but the small density of states at the Fermi surface
and the results for the other compounds suggest that this may just be
due to the well known gap underestimation of the PBE, and that a
quasiparticle gap may open up when more sophisticated
methods are used. Concerning the thermodynamical stability of these compounds,
CuCaO$_2$ is at 3\,meV/atom above the hull, while AuSrO$_2$ is at
33\,meV and AuBaO$_2$ at 31\,meV.

\subsection{Transition metals}

\begin{figure}
  \centering
  \begin{tabular}{cc}
    {\bf (a)} & {\bf (b)} \\  
    \includegraphics[width=.49\columnwidth,angle=0]{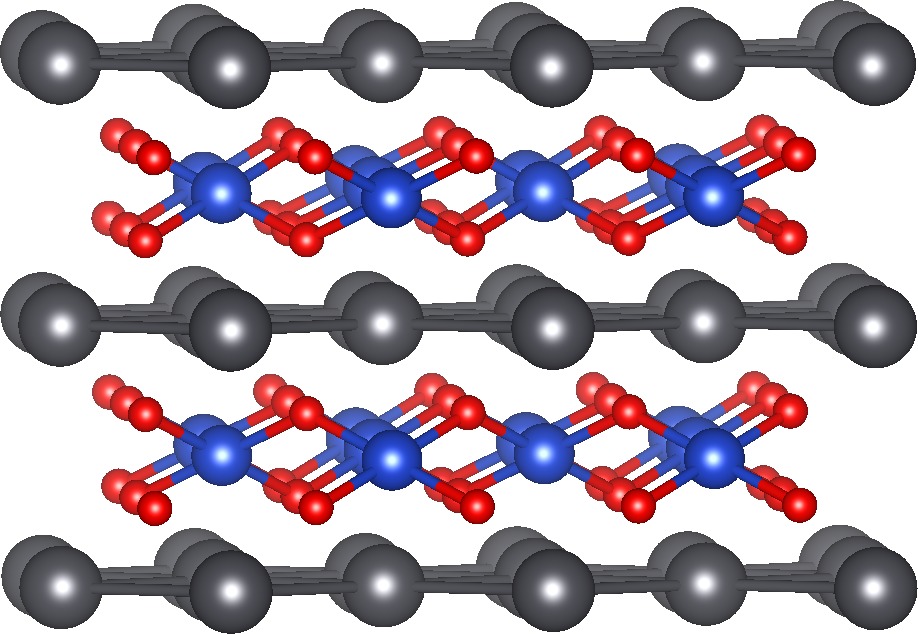} &
    \includegraphics[width=.44\columnwidth,angle=0]{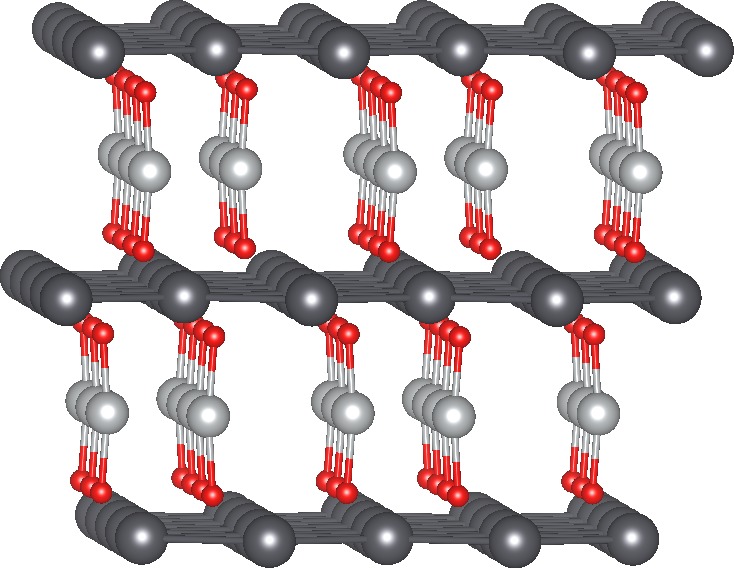}
  \end{tabular}

  {\bf (c)}

  \includegraphics[width=.49\columnwidth,angle=0]{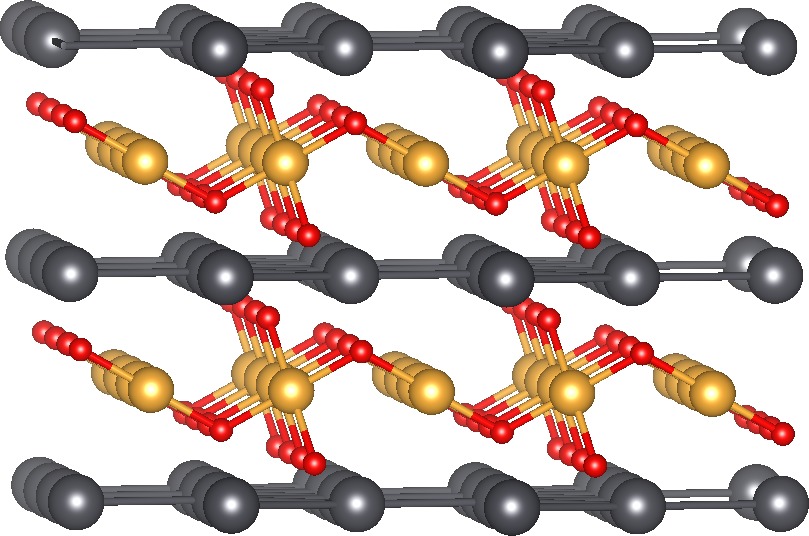}
  \caption{Structures of (a)~CuPbO$_2$ (space group 74), (b)~AgPbO$_2$ (space
    group 12), and (c)~AuPbO$_2$ (space group 2).}
  \label{fig:XPbO2}
\end{figure}

\paragraph{Vanadium}
The monoclinic structure (space group 12) of CuVO$_2$ is very similar
to the one of AgPbO$_2$ (see Fig.~\ref{fig:XPbO2}). This compound is
has a gap of 1.0\,eV and is 11\,meV/atom above the convex hull.

\paragraph{Manganese}
The oxide phases containing Mn (space group 12) can be seen as
distorted delafossites, where the O atoms are slightly displaced from
their symmetry positions (CuMnO$_2$, AuMnO$_2$), or with a different
stacking (AgMnO$_2$). The Au compound is metallic whereas the Ag and Cu have a
PBE gap of 0.4 and 0.1\,eV. 

%%%%%%%%%%%%%%%%%%%%%%%%%
\paragraph{Palladium, Platinum}

\begin{figure}[t]
  \includegraphics[width=.49\columnwidth,angle=0]{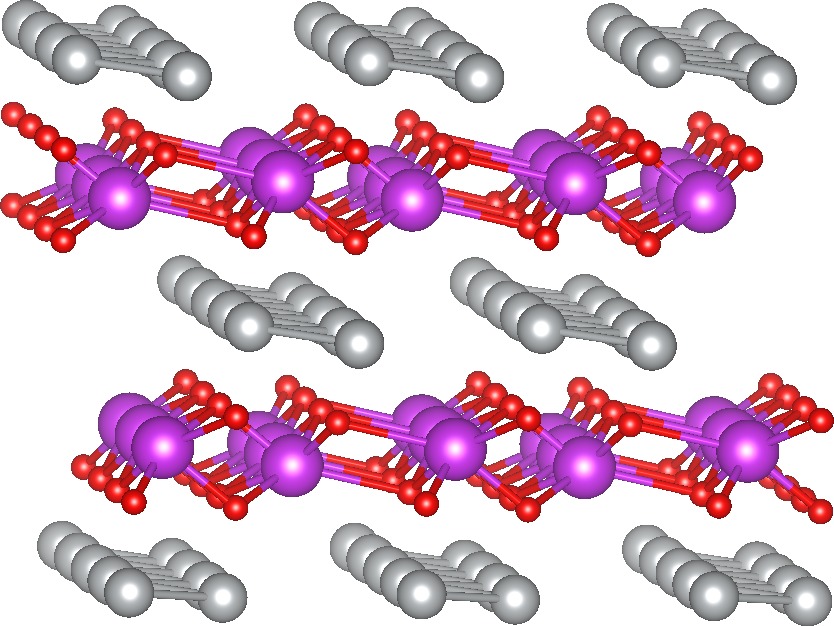}
  \caption{Distorted delafossite structure of (Cu,Ag,Au)BiO$_2$
    (compare with Fig.~\ref{fig:spg166}).}
  \label{fig:AgBiO2-02c-00_00014}
\end{figure}

The two Cu compounds with Pd and Pt crystallize in the orthorhombic
structure of (Ag,Au)LiO$_2$ (see right panel of
Fig.~\ref{fig:CuLiO2}). While the Pd compound is thermodynamically
stable, CuPtO$_2$ is 6\,meV/atom above the convex hull. Both are metallic. 
On the other hand, the two Ag phases have a monoclinic structure (space group 10)
similar to AuPbO$_2$ (see Fig.~\ref{fig:XPbO2}), but where the Ag
planes are not distorted (increasing the symmetry of the
system). AgPdO$_2$ is stable with a small PBE gap of 0.1\,eV. AgPtO$_2$ has an indirect gap of 0.3\,eV, lying 19\,meV above
the hull. Finally, AuPdO$_2$ it is a metallic distorted delafossite
structure (space group 12) similar to the one of (Cu,Ag,Au)BiO$_2$
(see Fig.~\ref{fig:AgBiO2-02c-00_00014}). 

\begin{figure}
  \centering
  \includegraphics[width=.49\columnwidth,angle=0]{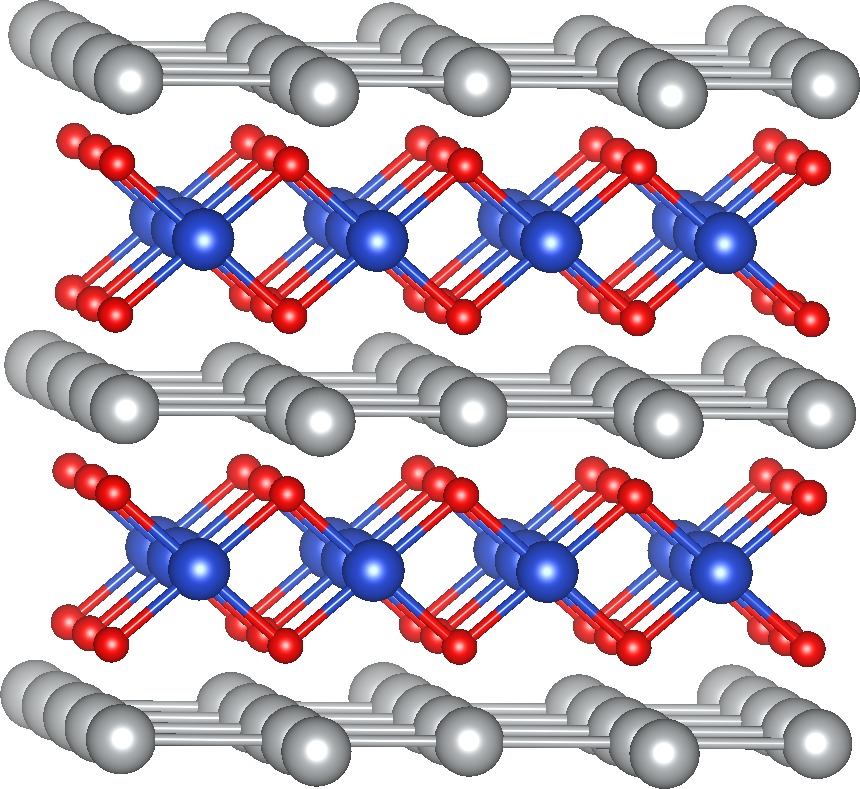}
  \caption{Structure of CuAgO$_2$ (space group 47).}
  \label{fig:CuAgO2}
\end{figure}

\paragraph{Copper, Silver, Gold}

CuAgO$_2$ has a orthorhombic lattice characterized by square flat
layers of Ag intercalated with ribbons of CuO$_2$. The main difference
to the structure of CuPbO$_2$ (see Fig.~\ref{fig:XPbO2}) is that the
four O atoms are shared in this case with two Cu neighbors and not
with four. On the other hand, CuAuO$_2$ crystallizes in a monoclinic
structure (space group 12) very similar to the Au delafossite structure,
with Au hexagonal flat planes separated by CuO$_2$ layers, but where
the O atoms are slightly displaced thereby reducing the symmetry from
trigonal to monoclinic. This is, in fact, the structure that can be
found in the databases for CuAgO$_2$. Finally, AgAuO$_2$ can again be
seen as a distorted Au delafossite, leading to a monoclinic phase with
space group 14. While CuAgO$_2$ and CuAuO$_2$ are metallic, AgAuO$_2$
is an indirect gap semiconductor with a PBE gap of 0.6\,eV.

\begin{figure}
  \centering
  \begin{tabular}{cc}
    {\bf (a)} & {\bf (b)} \\
    \includegraphics[width=.49\columnwidth,angle=0]{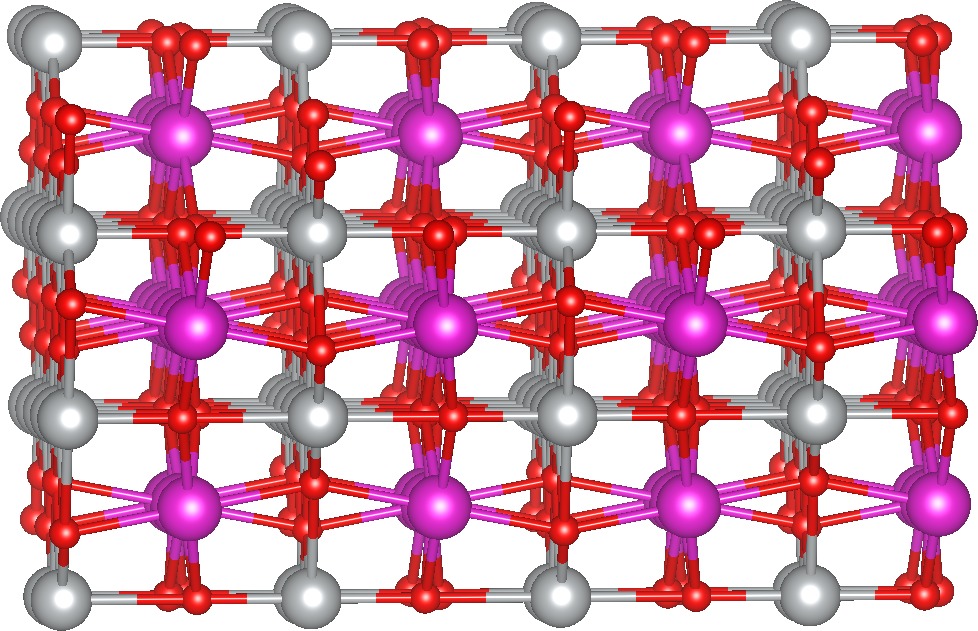} &
    \includegraphics[width=.49\columnwidth,angle=0]{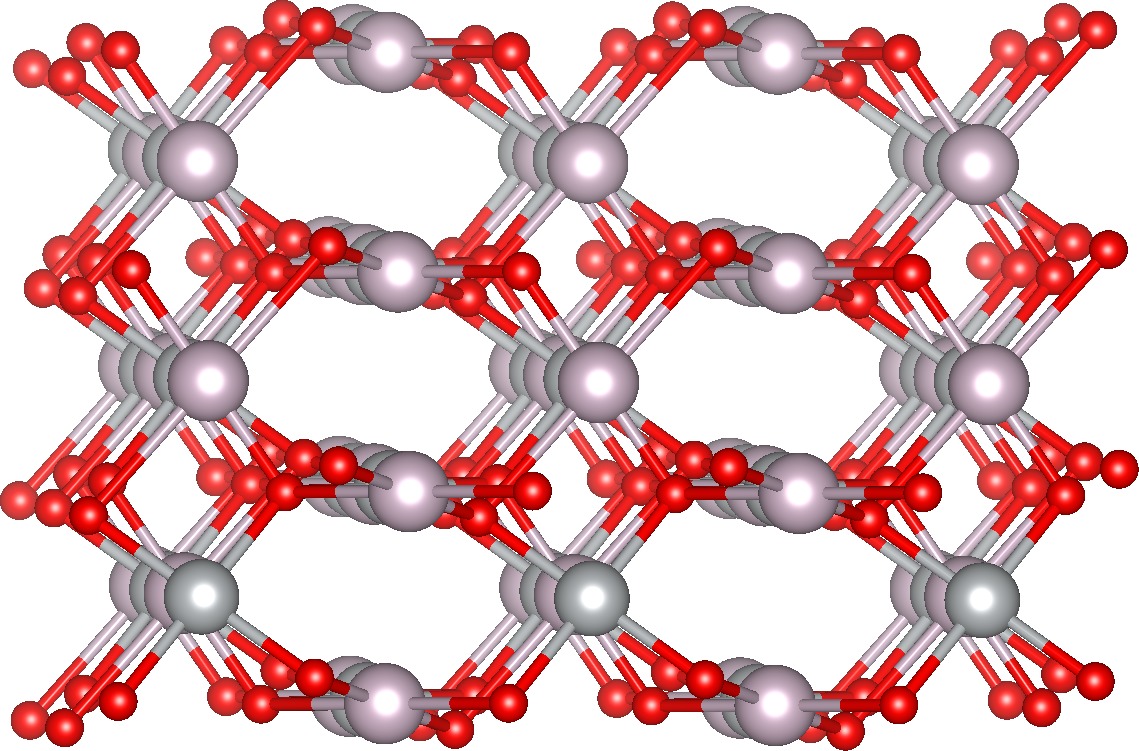}
  \end{tabular}
  \caption{Structure of (a)~AgCdO$_2$ (space group 12) and
    (b)~AgHgO$_2$ (space group 2).}
  \label{fig:AgCdO2}
\end{figure}

\paragraph{Cadmium, Mercury}

CuHgO$_2$ has the same monoclinic structure as CuAuO$_2$, and so it
can be seen as a deformed Hg delafossite. It is a metal and it lies 43\,meV/atom
above the convex hull of thermodynamic stability. On the other hand,
AgCdO$_2$ crystallizes in a deformed version (see
Fig.~\ref{fig:AgCdO2}) of the tetragonal structure of, e.g., CuCaO$_2$
(see Fig.~\ref{fig:CuCaO2}). This is a metal, 23\,meV/atom
above the hull. The next compound, AgHgO$_2$ has as ground state a low
symmetry structure (space group 2, see right panel of
Fig.~\ref{fig:AgCdO2}) that can be seen as a small distortion of the
orthorhombic phase of (Ag,Au)LiO$_2$ (see right panel of
\ref{fig:CuLiO2}). This is a metal, 27\,meV above the hull. Finally,
AuCdO$_2$ has a similar crystal structure as AgHgO$_2$, also with
space group 2, but with the metal atoms forming lines of Au and of Cd
(i.e., they are not alternating). This is a quasi-direct gap
semiconductor with a PBE gap of 0.9\,eV.

%%%%%%%%%%%%%%%%%%%%%%%%%
\begin{figure}[t]
  \includegraphics[width=.49\columnwidth,angle=0]{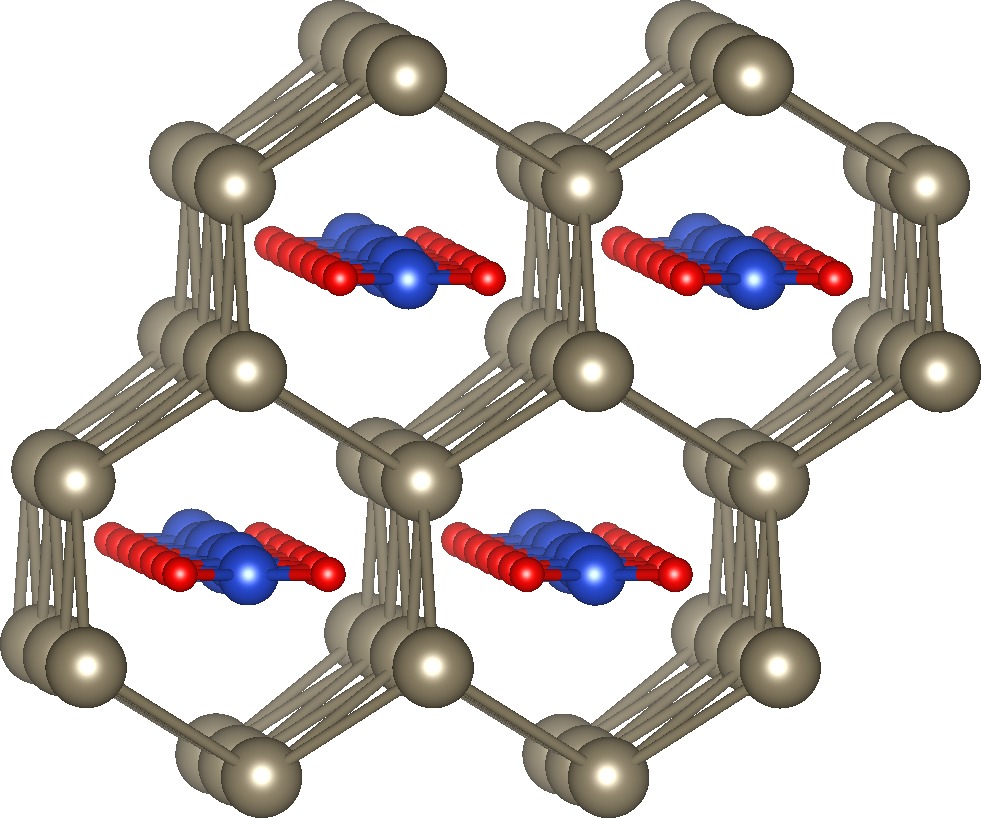}
  \caption{The structure of CuTlO$_2$ (space group 11).}
  \label{fig:spg11b}
\end{figure}

\paragraph{Thallium}
The ground-state of CuTlO$_2$, in contrast to the Ag and Au
compounds that crystallize in the delafossite structure, is a
monoclinic lattice (space group 11), characterized by hexagonal
channels made of Tl filled with flat stripes of CuO$_2$. This phase is
an indirect band-gap semiconductor with a PBE gap of 0.4\,eV. A (metallic)
delafossite structure is also present as a meta-stable phase, around
14\,meV/atom higher than the ground-state.

%%%%%%%%%%%%%%%%%%%%%%%%%

\paragraph{Lead}
(Cu,Ag,Au)PbO$_2$ oxides are characterized by Pb planes separated by
(Cu,Ag,Au)O$_2$ layers (see Fig.~\ref{fig:XPbO2}). For CuPbO$_2$, Pb
forms a square lattice while each Cu shares four O atoms with four Cu
neighbors, while for the Ag compound, Pb forms a hexagonal lattice
with isolated AgO$_2$ units in between. Finally, the AuPbO$_2$
structure is somewhat intermediate between the Cu and Ag compounds:
the Pb layers form a distorted hexagonal lattice and each Au has two O
atoms and shares an extra two. The Cu phase (space group 74) is
metallic and 43\,meV above the convex hull, the Ag compound (space
group 12) is a 0.3\,eV semiconductor and 32 meV above the hull, and finally Au
forms a semiconducting phase (space group 2) with an indirect PBE gap
of 0.4\,eV and it is 15\,meV above the convex hull.

%%%%%%%%%%%%%%%%%%%%%%%%%

\paragraph{Bismuth}

The ternary oxides with Bi are indirect-gap semiconducting phases with
PBE gaps of 1.0\,eV (Cu), 1.4\,eV (Ag), and 1.3\,eV (Au). Their
monoclinic space group 11 corresponds to an atomic arrangement that
can be seen as a distorted delafossite, with a dimerization of the
chains of Ag and BiO$_2$ (see
Fig.~\ref{fig:AgBiO2-02c-00_00014}). The delafossite structure
remains, however, a meta-stable phase of this composition, around
40\,meV/atom higher than the ground-state. Although not strictly
stable thermodynamically, these structures are remarkably close to the
convex hull at 8 meV/atom (Cu), 3 meV/atom (Ag), and 40 meV/atom (Au).

\subsection{Halogens}

\begin{figure}[t]
  \centering
  \includegraphics[width=.49\columnwidth,angle=0]{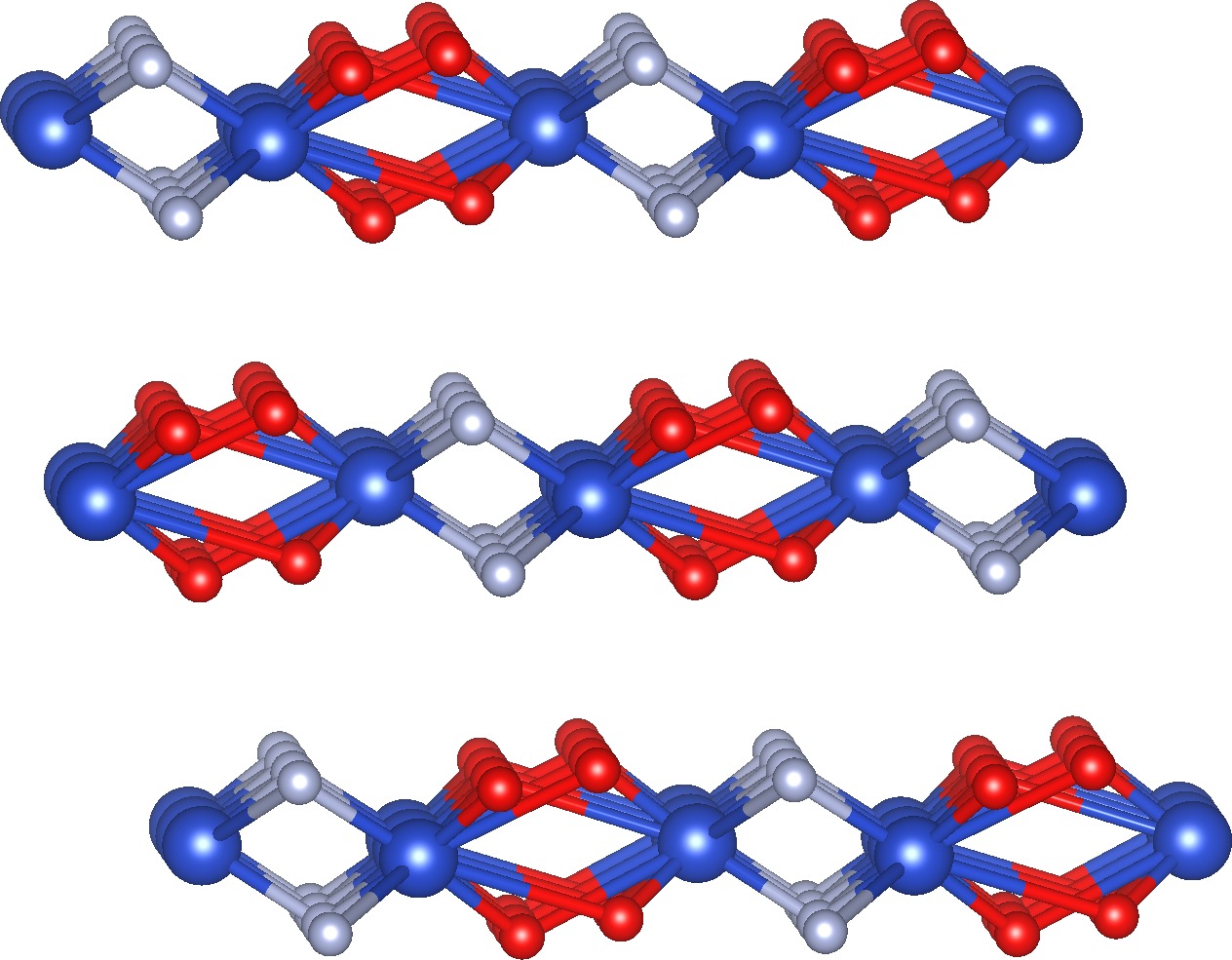}
  \caption{The structure of (Cu,Ag,Au)FO$_2$ (space group 4).}
  \label{fig:CuFO2}
\end{figure}

\paragraph{Fluorine}

(Cu,Ag,Au)FO$_2$ compounds crystallize in a monoclinic lattice (space
group 4) characterized by (Cu,Ag,Au)FO$_2$ layers (see
Fig.~\ref{fig:CuFO2}). These are all semiconducting structures, with
PBE electronic gaps of 0.8\,eV (CuFO$_2$), 0.6\,eV (AgFO$_2$),
and 1.1\,eV (AuFO$_2$).

%\begin{figure}[t]
%  \centering
%  \begin{tabular}{cc}
%    {\bf (a)} & {\bf (b)} \\  
%    \includegraphics[width=.42\columnwidth,angle=0]{img/spg11c} &
%    \includegraphics[width=.42\columnwidth,angle=0]{img/spg15a}
%  \end{tabular}
%
%  {\bf (c)}
%
%  \includegraphics[width=.49\columnwidth,angle=0]{img/spg20a}
%  \caption{The structure of (a)~ (space group 11),
%    (b)~ (space group 15), and (c) (space group 20).}
%  \label{fig:XXXXO2}
%\end{figure}

\begin{figure}[t]
  \centering
  \begin{tabular}{cc}
    {\bf (a)} & {\bf (b)} \\
    \includegraphics[width=.45\columnwidth,angle=0]{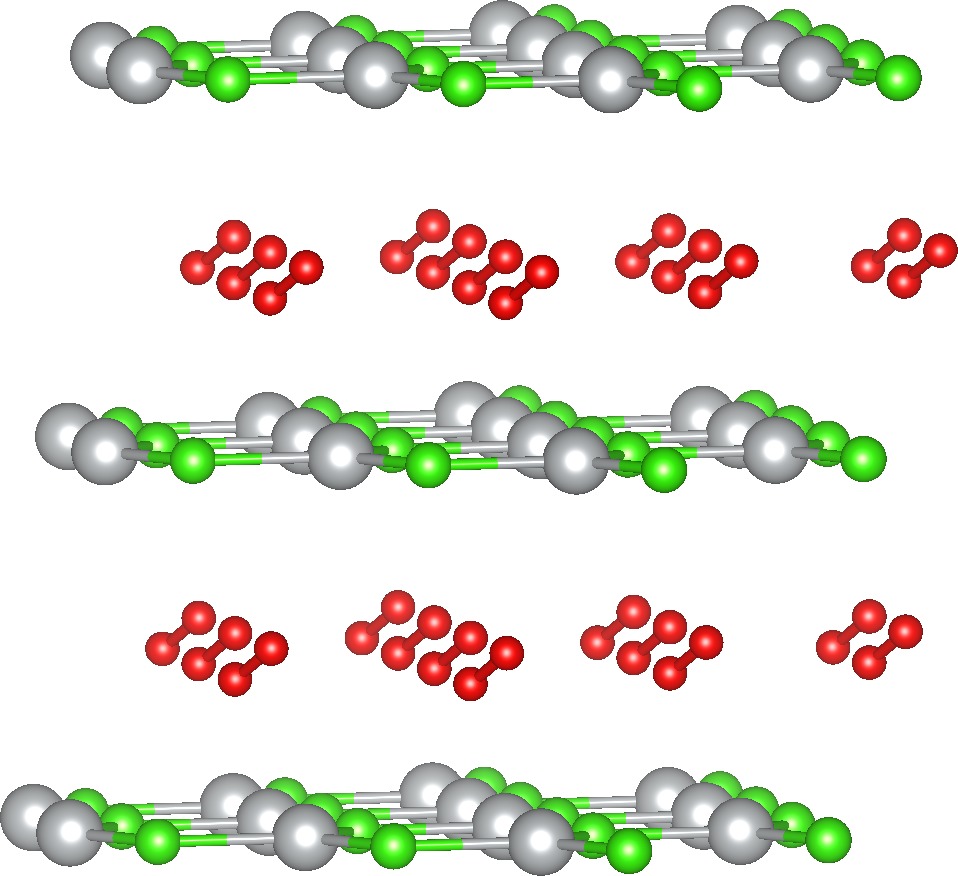} &
    \includegraphics[width=.45\columnwidth,angle=0]{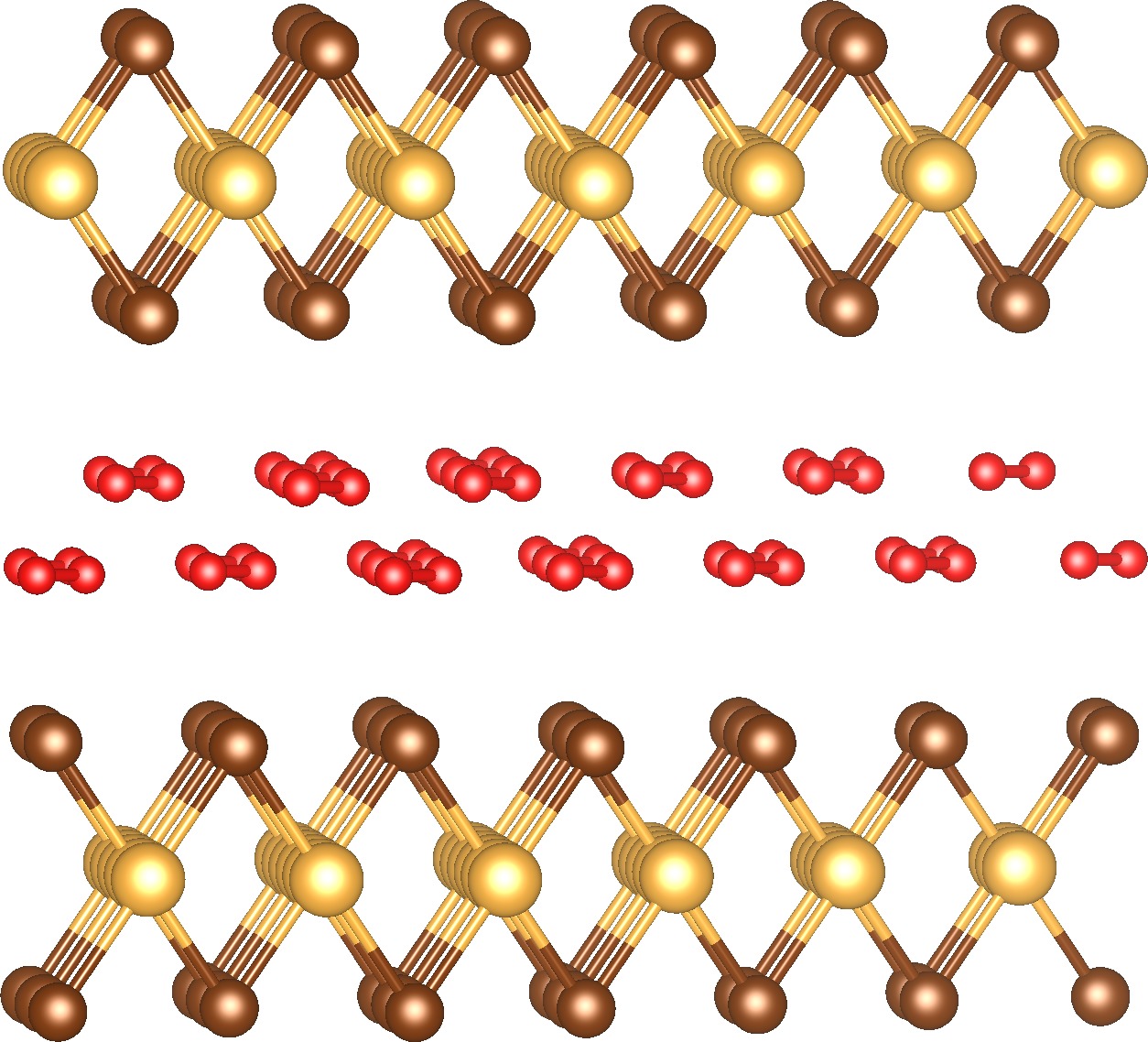}
  \end{tabular}

  {\bf (c)}

  \includegraphics[width=.79\columnwidth,angle=0]{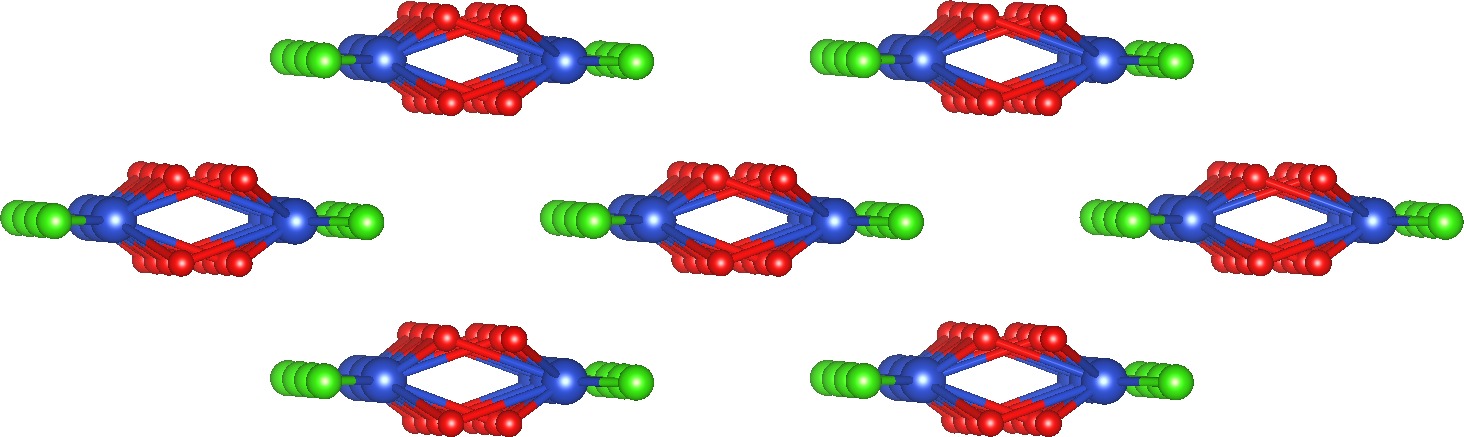}

  \caption{The structure of (a)~Ag(Cl,Br)O$_2$ (space group 1),
    (b)~Au(Cl,Br)O$_2$ (space group 15), and (c)~Cu(Cl,Br)O$_2$ (space
    group 1).}
  \label{fig:CuClO2}
\end{figure}

\paragraph{Chlorine, Bromine}

Cu(Cl,Br)O$_2$ compounds crystallize in a low symmetry triclinic
(space group 1) lattice (see panel c of Fig.~\ref{fig:CuClO2}). This
has some similarities to the monoclinic structure of (Cu,Ag,Au)FO$_2$
(see Fig.~\ref{fig:CuFO2}), in the sense that they share the same
CuO$_2$ subunits. However, in the latter structure each Cu shares four
F with four neighboring Cu atoms, while in the former each Cu only
shares two (Cl,Br) with two other Cu. These are thermodynamically
stable structures, that are indirect gap semiconductors with PBE gap
of 0.8 (CuClO$_2$) and 0.9 (CuBrO$_2$).

Ag(Cl,Br)O$_2$ structures are fundamentally different from all other
we found in our study. It is composed of flat hexagonal layers of
Ag(Cl,Br), separated by a layer of O$_2$ molecules (see panel a of
Fig.~\ref{fig:CuClO2}), forming a low-symmetry triclinic lattice
(space group 1). These turn out to be semiconductors with PBE gaps of 0.6\,eV
(AgClO$_2$) and 0.7\,eV (AgBrO$_2$).

Au(Cl,Br)O$_2$ compounds crystallize in a lattice that is composed of
Au(Cl,Br) layers intercalated with O$_2$ molecules (see panel b of
Fig.~\ref{fig:CuClO2}). However, in this case the Au(Cl,Br) are not
flat, but form zigzag stripes with each Au bonded to two (Cl,Br) and
vice-versa, leading to a monoclinic lattice (space group
15). AuClO$_2$ is a semiconductor with an indirect gap of 0.3\,eV,
while AuBrO$_2$ turns out to be a metal with the PBE.

\subsection{Others}

\begin{figure}[t]
  \centering
  \includegraphics[width=.49\columnwidth,angle=0]{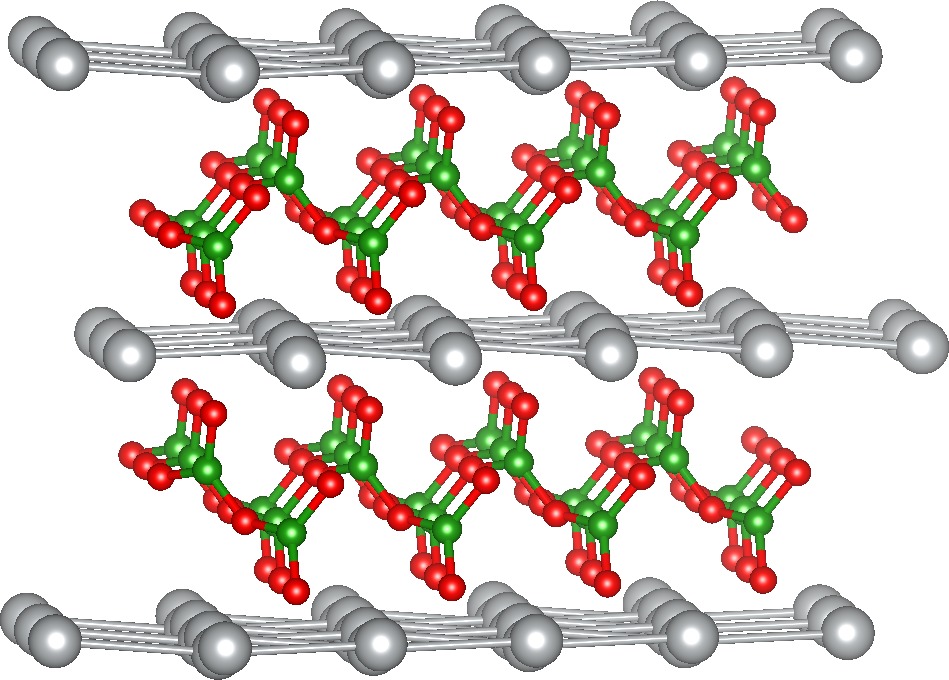}
  \caption{The structure of AgBO$_2$ (space group 9).}
  \label{fig:AgBO2}
\end{figure}

\paragraph{Boron}

AgBO$_2$ has a monoclinic structure (space group 9) composed by
slightly buckled Ag layers, separated by BO$_2$ chains, where each B
is connected to three O, two of which are shared with two adjacent B
atoms (see Fig.~\ref{fig:AgBO2}). Note that this is the same structure
we had previously found for CuBO$_2$~\cite{MRS_comm}. AgBO$_2$ is
33\,meV/atom above the hull, and it is a semiconductor with a PBE
quasi-direct gap of 1.4 eV.

\begin{figure}[t]
  \centering
  \begin{tabular}{cc}
    {\bf (a)} & {\bf (b)} \\  
    \includegraphics[width=.47\columnwidth,angle=0]{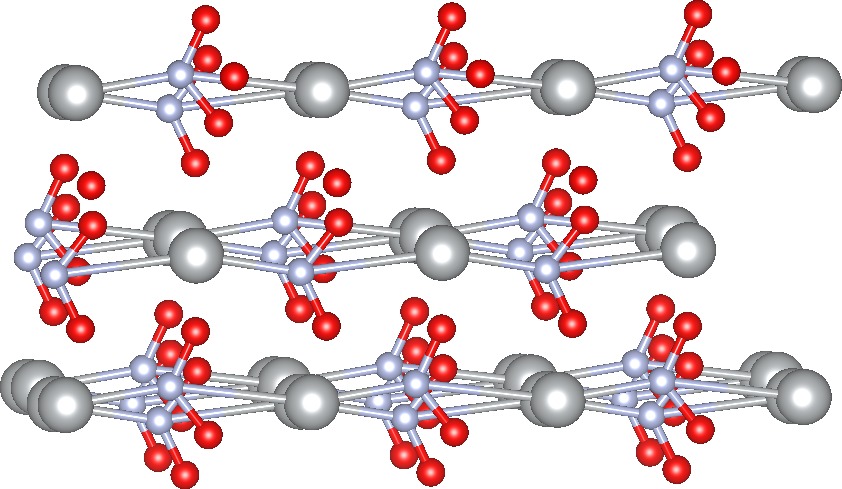} &
    \includegraphics[width=.51\columnwidth,angle=0]{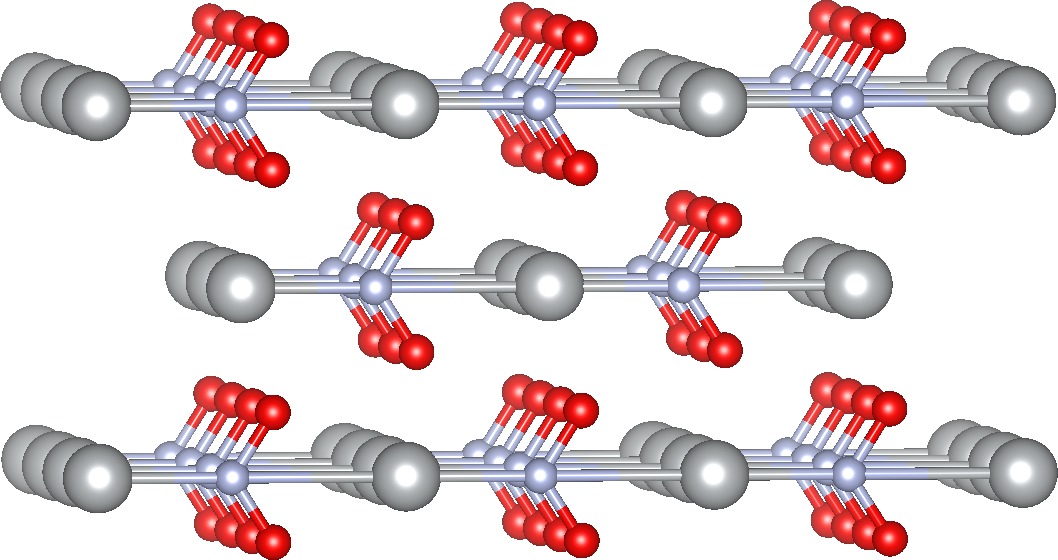}
  \end{tabular}
  \caption{Two phases of AgNO$_2$, (a)~the ground-state monoclinic
    (space group 5)and an orthorhombic (space group 44) structures.}
  \label{fig:AgNO2}
\end{figure}

\paragraph{Nitrogen}
The lowest energy phase of AgNO$_2$ that we found
is a monoclinic structure (space group 5), lying 37\,meV/atom above
the hull. It is composed of lines of alternating Ag and NO$_2$. This
is a deformation of the orthorhombic structure (space group 44, see
Fig.~\ref{fig:AgNO2}) that can be found in the databases, and that we
found to be 5\,meV/atom above the ground state. Both structures are
indirect gap semiconductors, with PBE gaps of 1.4\,eV (monoclinic)
and 1.8\,eV (orthorhombic).

\begin{figure}[t]
  \centering
  \includegraphics[width=.47\columnwidth,angle=0]{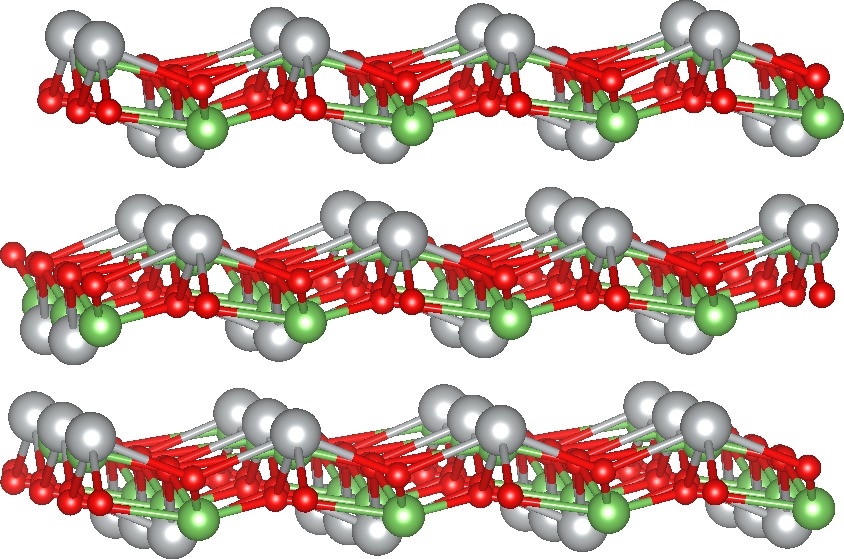}
  \caption{The structure of AgAsO$_2$ (space group 5).}
  \label{fig:AgAsO2}
\end{figure}

\paragraph{Arsenic}
Also for AgAsO$_2$ we found a monoclinic structure (space group 5,
see Fig.~\ref{fig:AgAsO2}), composed of corrugated planes of
AgAsO$_2$. This phase is 38\,meV/atom above the hull, with a PBE
indirect gap of 2.4\,eV.

\section{Conclusion}

In summary, we performed structural prediction runs for 183 oxide
phases containing Cu, Ag, and Au. From our runs we predict that there
are  81 thermodynamically stable or quasi-stable (within 20\,meV)
compositions, out of which only 36 are included in available
databases. These numbers should be compared to the 3 systems that were
found in a previous study using high-throughput techniques combined
with machine learning. Nevertheless, we believe that structural
prediction should not be seen as a competitor of more traditional
high-throughput techniques. In fact, both methods complement each
other, and both should be used in synergy to speed up the experimental
process of materials discovery. We tried to do the first steps in this
direction, by combining structural prediction and a search based on
prototype structures.

The subset of oxides that we explored include delafossite CuAlO$_2$,
the first delafossite p-type transparent conductive oxide. By calculating the
band gaps and hole effective masses of the new stable compounds that
we identified, we could reveal some potential candidates to outperform
CuAlO$_2$ as p-type transparent conductor. These few compounds deserve in
our mind further consideration, both from an experimental point of
view (e.g., synthesis, structural and electronic characterization) and
a theoretical point of view (e.g., study of dopability, more accurate
band structure and transport calculations).

Certainly, structural prediction is complex and numerically
expensive. However, with our results we aim at demonstrating that,
using modern supercomputers, it is possible to use such techniques to
investigate a large number of chemical compositions, paving the way
for a large scale search for new materials.

\begin{acknowledgement}
SB and MALM acknowledge support from the French ANR
(ANR-12-BS04-0001-02). Financial support provided by the Swiss
National Science Foundation is gratefully acknowledged. Computational
resources were provided by GENCI (project x2013096017) in France and
the PRACE-3IP project (FP7 RI-312763) resource Archer based in
Scotland at the University of Edinburgh.
\end{acknowledgement}

\begin{suppinfo}
Crystallographic Information Files for the crystal structures studied in this work. 
\end{suppinfo}

\bibliography{biblio}

\end{document}